\documentclass{jfm}

\usepackage{xr}
\usepackage{mathtools}
\usepackage{lmodern}
\usepackage{gensymb}
\usepackage{multirow}
\usepackage{url}
\usepackage{setspace}
\usepackage{natbib}
\usepackage[percent]{overpic}
\usepackage{amssymb}
\usepackage{amsmath}
\usepackage{mathrsfs}
\usepackage{mathtools}
\usepackage{cancel}
\usepackage{soul}
\usepackage{float}
\usepackage{textcomp}
\usepackage{listings}
\usepackage{psfrag}
\usepackage{placeins}
\usepackage{color}
\usepackage{epsfig}
\usepackage{bm}
\usepackage{lscape}
\usepackage[usenames,dvipsnames,table]{xcolor}
\usepackage{tikz}
\usetikzlibrary{positioning,arrows}
\usetikzlibrary{decorations.pathreplacing}
\usepackage{pstricks}
\usepackage[toc,page]{appendix}
\usepackage{upgreek}
\usepackage{caption}
\usepackage{multirow}
\usepackage[mathlines]{lineno}
\usepackage{scalerel}
\allowdisplaybreaks
\usepackage{anyfontsize}
\usepackage{mathrsfs}

\usepackage{emptypage}
\usepackage{natbib}
\usepackage{graphics}
\usepackage{graphicx}
\usepackage{amssymb}
\usepackage{amsmath}
\usepackage{epsfig,multirow}
\usepackage{bm,xparse}
\usepackage[intoc]{nomencl}
\usepackage[breaklinks=true]{hyperref}
\usepackage[usenames,dvipsnames,table]{xcolor}
\usepackage{tikz,datetime}
\usepackage{verbatim}
\usepackage{float}
\usepackage{subcaption}
\usepackage{epigraph}
\usepackage[shortlabels]{enumitem}
\usepackage{pgfplots}
\usetikzlibrary{spy}
\usetikzlibrary{decorations.markings}
\usepackage{cancel}
\usepackage{csquotes}
\usepackage{tikz,tikzscale,pgfplots,etoolbox,pgfplotstable}
\usepgfplotslibrary{groupplots}
\usepgfplotslibrary{colorbrewer}
\usepgfplotslibrary{colormaps}
\usepgfplotslibrary{fillbetween}
\usetikzlibrary{patterns}
\usetikzlibrary{arrows,calc,positioning,shapes,arrows,external,cd}
\AtBeginEnvironment{tikzcd}{\tikzexternaldisable}
\AtEndEnvironment{tikzcd}{\tikzexternalenable}

\newcommand{\PLH}{{\mkern-2mu\times\mkern-2mu}}

\newcommand{\avertime}[1]{\overline{#1}}

\newcommand{\aversymmetry}[1]{\left[#1\right]_s}
\newcommand{\averspace}[1]{\langle #1 \rangle}
\newcommand{\averglobal}[1]{\left[#1\right]_g}

\newcommand{\xs}{x_s}
\newcommand{\zs}{z_s}

\makeatletter
\def\ps@myheadings{%
    \let\@oddfoot\@empty\let\@evenfoot\@empty
    \def\@evenhead{\thepage\hfil\slshape\leftmark}%
    \def\@oddhead{{\slshape\rightmark}\hfil\thepage}%
    \let\@mkboth\@gobbletwo
    \let\sectionmark\@gobble
    \let\subsectionmark\@gobble
    }
\renewcommand\maketitle{\par
  \begingroup
    \renewcommand\thefootnote{\@fnsymbol\c@footnote}%
    \def\@makefnmark{\rlap{\@textsuperscript{\normalfont\@thefnmark}}}%
    \long\def\@makefntext##1{\parindent 1em\noindent
            \hb@xt@1.8em{%
                \hss\@textsuperscript{\normalfont\@thefnmark}}##1}%
    \if@twocolumn
      \ifnum \col@number=\@ne
        \@maketitle
      \else
        \twocolumn[\@maketitle]%
      \fi
    \else
      \newpage
      \global\@topnum\z@   
      \@maketitle
    \fi
    \thispagestyle{plain}\@thanks
  \endgroup
  \setcounter{footnote}{0}%
}
\makeatother

\title[Reduction of friction drag by passively rotating discs]{Reduction of turbulent skin-friction drag by passively rotating discs}

\author[Paolo Olivucci, Daniel J. Wise, Pierre Ricco]
{Paolo Olivucci$^1$, Daniel J. Wise$^2$ and Pierre Ricco$^1$\thanks{Email address for correspondence: p.ricco@sheffield.ac.uk}}
\affiliation{$^1$Department of Mechanical Engineering, The University of Sheffield, S1 3JD Sheffield, United Kingdom\\[\affilskip]
$^2$Department of Fluid Dynamics, A*Star Institute of High Performance Computing, Singapore}
\pubyear{}
\volume{}
\pagerange{}

\begin{document}

\maketitle


\begin{abstract}
A turbulent channel flow modified by the motion of discs that are free to rotate under the action of wall turbulence is studied numerically. The Navier-Stokes equations are coupled nonlinearly with the dynamical equation of the disc motion, which synthesizes the fluid-flow boundary conditions and is driven by the torque exerted by the wall-shear stress. We consider discs that are fully exposed to the fluid and discs for which only half of the surface interfaces the fluid. The disc motion is thwarted by the fluid torque in the housing cavity and by the torque of the ball bearing that supports the disc. For the full discs, no drag reduction occurs because of the small angular velocities. The most energetic disc response occurs for disc diameters that are comparable with the spanwise spacing of the low-speed streaks. A perturbation analysis for small disc-tip velocities reveals that the two-way nonlinear coupling has an intense attenuating effect on the disc response. The reduced-order results show excellent agreement with the nonlinear results for large diameters. The half discs rotate with a finite angular velocity, leading to large reduction of the turbulence activity and of the skin-friction drag over the spinning portion of the discs, while the maximum drag reduction over the entire walls is 5.6\%. The dependence of the drag reduction on the wall-slip velocity and the spatial distribution of the wall-shear stress qualitatively match results based on the only available experimental data.
\end{abstract}


\section{Introduction}
\label{sec:introduction}

Turbulent skin-friction drag reduction is a subject of great interest in fluid mechanics research, due to the potential to produce significant reductions in fuel consumption and carbon emissions in many industrial scenarios. Around $60\%$ of the aerodynamic drag of a typical airliner in cruise conditions is due to frictional drag, the other components being form drag and induced drag \citep{leschziner-etal-2011}. In the case of a typical slow merchant ship, such as a tanker or a container ship, around $70\%$ of the total hull resistance is frictional drag, the remainder being form drag and wave drag \citep{larsson-raven-2010}. For the aerospace industry, it has been estimated that a $1\%$ reduction in skin-friction drag for a long-range commercial aircraft would decrease fuel consumption by approximately 0.45\%. For a typical airline this change is equivalent to a reduction in CO$_2$ emissions of 5.4 million equivalent tonnes per year and an annual reduction in operating costs of 0.2\% \citep{reneaux-2004}.

At the moderate Reynolds numbers considered in this study, the near-wall dynamics is responsible for the largest contribution to the skin-friction drag \citep{quadrio-2011,degiovanetti-etal-2016}. Wall-based drag reduction techniques therefore often aim at altering the structure of the near-wall turbulence, producing an attenuation of the ejections and sweeps that promote the transport of high-momentum fluid in the near-wall region and lead to an increased skin-friction drag.

Wall-based flow control methods for drag reduction can be broadly classified as active or passive. In active methods, the flow is often altered by wall-mounted devices that rely on an external power supply to operate. For such methods, the power required to drive the control devices must be accounted for because the net power consumption represents a critical performance metric. Passive methods instead consist of modifications of the wall surface geometry \citep{min-kim-2004,garcia-jimenez-2011} or compliant surfaces where some regions of the wall are allowed to move under the action of the flow \citep{choi-etal-1997}. Passive methods do not necessitate an external power input, although the interaction between the fluid and the complex geometry may constitute an additional source of friction.

\cite{jung-1992} first introduced in-plane wall motion as a drag-reduction technique, achieving an attenuation of the turbulent skin-friction drag by oscillating the wall in the cross-flow direction. The oscillating wall technique has proven capable of generating drag-reduction margins up to $40\%$ \citep{quadrio-ricco-2004}, while its generalizations in the form of steady and travelling waves of spanwise velocity at the wall have also led to drag-reduction levels as high as $45\%$ at low Reynolds numbers \citep{quadrio-ricco-viotti-2009}. The mechanism by which the spanwise wall motion suppresses the turbulent activity and leads to drag reduction has been the subject of several investigations \citep{choi-2002,quadrio-2011,skote-2011,ricco-etal-2012,blesbois-etal-2013}. An optimal oscillation period of about $T^+=100$ (the superscript ``+'' denoting scaling in viscous units) was found by \cite{quadrio-ricco-2004} to produce the maximum drag reduction because it matches a characteristic temporal scale of the near-wall turbulent structures and an optimal thickness of the generated Stokes layer.

Spinning circular actuators that are flush-mounted on the wall have also been demonstrated numerically to lead to drag-reduction levels up to $23\%$ and to net power savings up to $10\%$ \citep{ricco-hahn-2013,wise-ricco-2014}, where the latter is computed by subtracting the power spent to move the discs from the power saved because of drag reduction. The discs offer a few practical advantages with respect to wall oscillations and travelling waves, i.e., the local character that avoids the motion of the whole solid wall and the wider range of diameters leading to drag-reduction margins that are comparable to the maximum value \citep{ricco-hahn-2013,wise-alvarenga-ricco-2014}. \cite{wise-alvarenga-ricco-2014} demonstrated that rotating annular actuators can deliver similar levels of drag reduction while requiring up to $20\%$ less driving power. The ring actuators can be combined efficiently with other passive and active techniques \citep{olivucci-ricco-2019}, attaining reductions of the skin-friction drag as large as $27\%$.

In the present work, we carry out direct numerical simulations of a turbulent channel flow over flush-mounted discs that are free to move under the action of the turbulent wall-shear stress. We consider the full numerical modelling of the rotating actuators, taking into account the coupled dynamics of the turbulent flow and the discs, and the frictional losses occurring in a realistically designed disc housing.
 
Modelling of flow-control actuators is critical because it gives estimates of the power expenditure that are closer to those in real-world settings. This approach also allows studying the two-way coupling between the fluid mechanics and the actuators, which leads to a more accurate prediction of the flow physics and therefore of the drag-reduction performace. In the case of the discs, the two-way coupling is able to capture the detailed physics of the stress and torque fluctuations to which a disc rotating under a turbulent flow is subjected. \cite{jozsa-etal-2019} simulated the coupled fluid-body dynamics of passive, shear-stress-driven cylindrical actuators as a compliant-wall technique able to generate a finite in-plane wall velocity and give drag reduction. \cite{mahfoze-etal-2018} simulated suction and blowing in a boundary layer and used an empirical model of the frictional losses in the electromagnetic speaker used to generate the wall-normal transpiration. To the authors' knowledge, there have been no attempts to model the coupled dynamics between a fluid and in-plane flow-control actuators, and to estimate the mechanical losses realistically.

The dynamics of freely-rotating full discs and half discs under the shearing action of the wall turbulence is studied. The first case serves the purpose of analyzing the unsteady response of a full disc to the wall turbulence and to verify the numerical implementation of the two-way coupled dynamics between the fluid and the disc, thereby allowing a detailed study of the fluid-structure interaction problem. The half-disc case is subsequently investigated with the primary objective to explore the potential for turbulent skin-friction drag reduction. The half discs are realized by preventing the left spanwise halves of freely rotating discs to interact with the flow, thereby producing a finite average shear-stress torque on the disc surfaces. The half discs thus rotate with a finite angular velocity and introduce a slip velocity at the wall, thereby affecting the wall-shear stress. 

\begin{figure}
\begin{subfigure}[l]{.43\textwidth}
\caption{}
	\includegraphics[scale=.35]{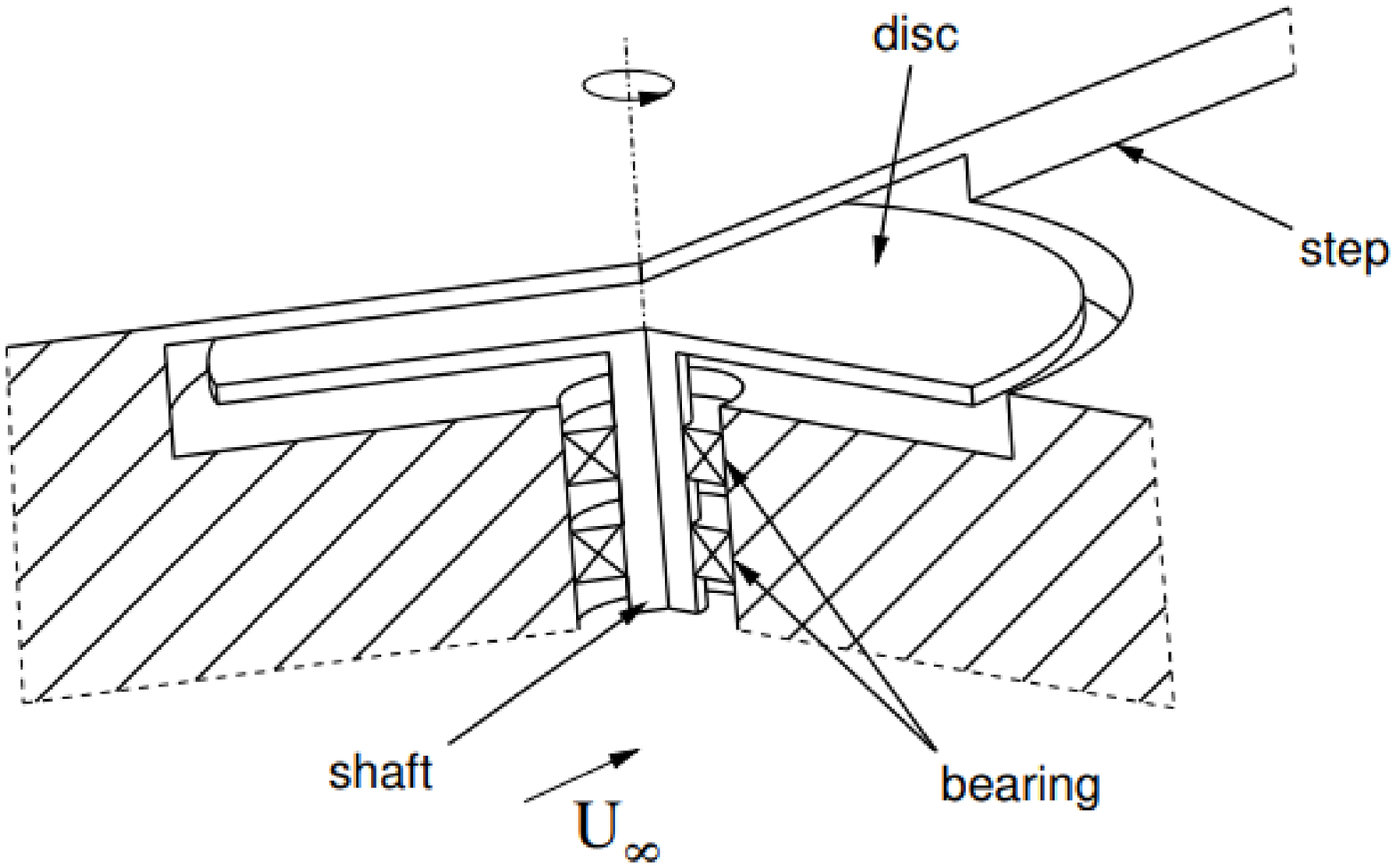}
\end{subfigure}
\begin{subfigure}[l]{.40\textwidth}
\caption{}
	\includegraphics[scale=.42]{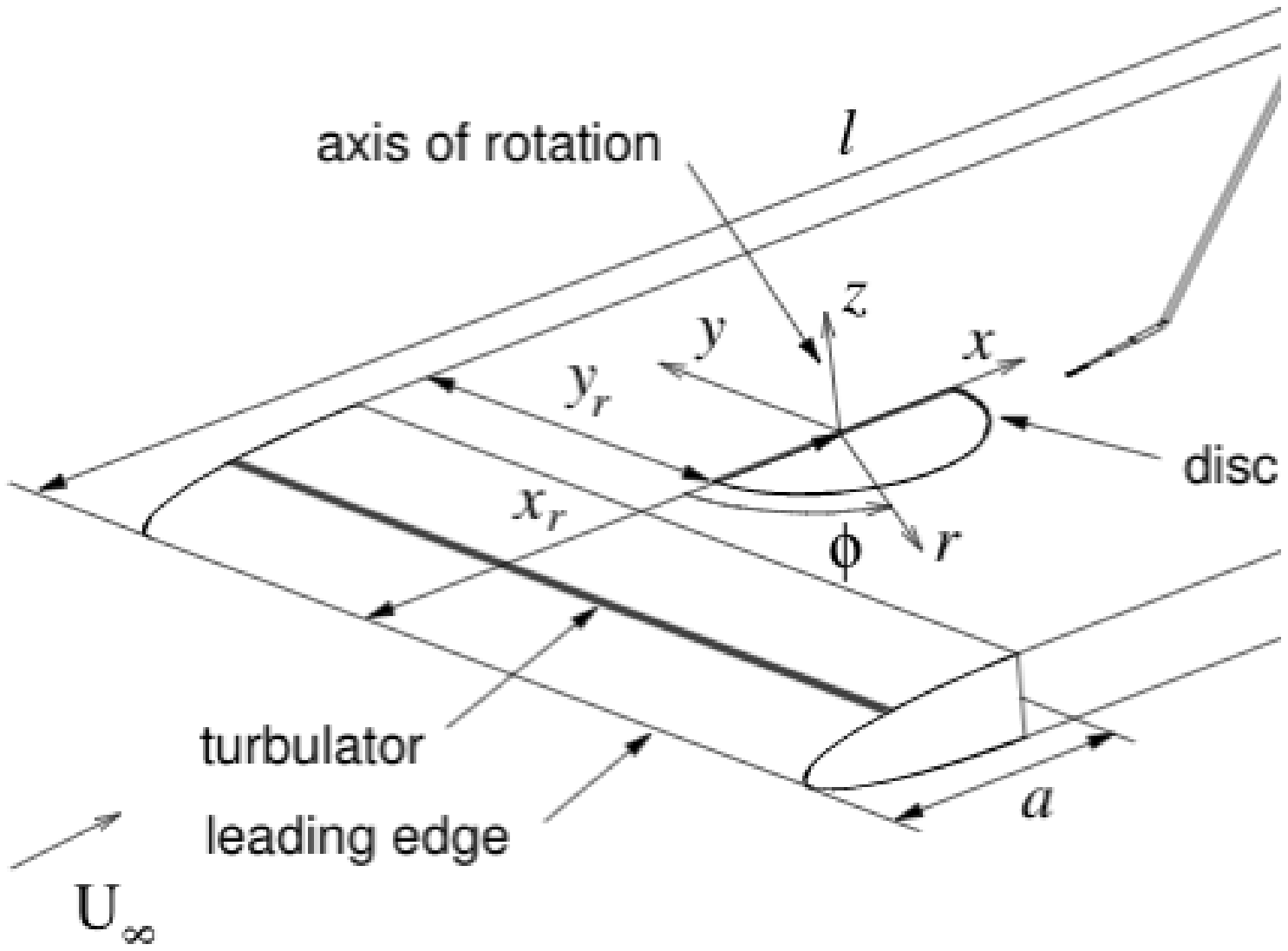}
\end{subfigure}
\vspace{0.5cm}
\caption{Experimental apparatus of \cite{koch-kozulovic-2013} and \cite{koch-kozulovic-2014}. (a) Schematic of the disc and its supporting housing system. (b) Schematic of the wind-tunnel wall, fitted with a half disc. Taken from \cite{koch-kozulovic-2013} with permission from American Society of Mechanical Engineers.} 
\label{fig:setups}
\end{figure}

Our study is also motivated by the experimental boundary-layer study of two configurations of half discs by \cite{koch-kozulovic-2013} (KK13) and \cite{koch-kozulovic-2014} (KK14), shown in figure \ref{fig:setups}. The wall-shear stress was not measured on the disc surface, but it was estimated by modifying a correlation for computing the skin-friction coefficient in uncontrolled boundary layers. Koch and Kozulovic used the local streamwise slip velocity generated by the disc in the correlation, thus predicting reductions of skin-friction drag up to $17\%$ on the disc surface. The procedure adopted by KK13 and KK14 is detailed in the Supplementary Material \ref{app:kk}.

The numerical and statistical procedures are presented in \S\ref{sec:numerics}. Section \ref{sec:model} describes the dynamics of the actuators and the modelling of the torques acting on the discs. In \S\ref{sec:results-full}, we study the passive response of full freely-rotating discs to the turbulent flow. In \S\ref{sec:results-half}, the simulations of arrays of half discs are discussed and compared to the experimental results of KK13 and KK14. Section \ref{sec:conclusions} presents the conclusions of the study. 

\section{Numerical procedures}
\label{sec:numerics}

In this section, we present the flow system and the numerical procedures. An incompressible turbulent flow between parallel flat walls fitted with spinning discs is investigated. The flat discs, flush-mounted on the walls, are free to rotate under the viscous action of the turbulent channel flow. The Cartesian coordinates describing the system are the streamwise coordinate $x^*$, the wall-normal coordinate $y^*$, and the spanwise coordinate  $z^*$ (the $*$ superscript indicates a dimensional quantity). The respective velocity components in these directions are $u^*$, $v^*$, and $w^*$. The channel walls are separated by a distance $L_y^*=2h^*$ and the flow is driven by a streamwise pressure gradient. A constant flow rate per unit wall-area $Q^*$ is imposed. The bulk velocity is $U^*_b=Q^*/2h^*$. 

The fluid motion obeys the incompressible continuity and Navier-Stokes equations (NSE),
\begin{align}
\nabla \cdot \mathbf{u} & =0,
\label{eq:continuity-cartesian-vector} \\
\frac{\p \mathbf{u}}{\p t}+\left(\mathbf{u}\cdot \nabla \right)\mathbf{u} & =-\nabla p+\frac{1}{Re_p}\nabla^2 \mathbf{u}. 
\label{eq:navier-stokes-cartesian-vector} 
\end{align}
Quantities are non-dimensionalized by the channel half-height $h^*$, the density of the fluid $\rho^*$, and the centreline velocity of the laminar Poiseuille flow at the same flow rate, defined as $U_p^*=3Q^*/4h^*$. Quantities expressed in these outer units are written without any symbol. All flows are at a Reynolds number $Re_p = U_p^* h^* / \nu^* = 4200$, where $\nu^*$ is the kinematic viscosity of the fluid. 

\begin{figure}
\centering
\begin{subfigure}[l]{.8\textwidth}
\centering
    \includegraphics[width=\textwidth]{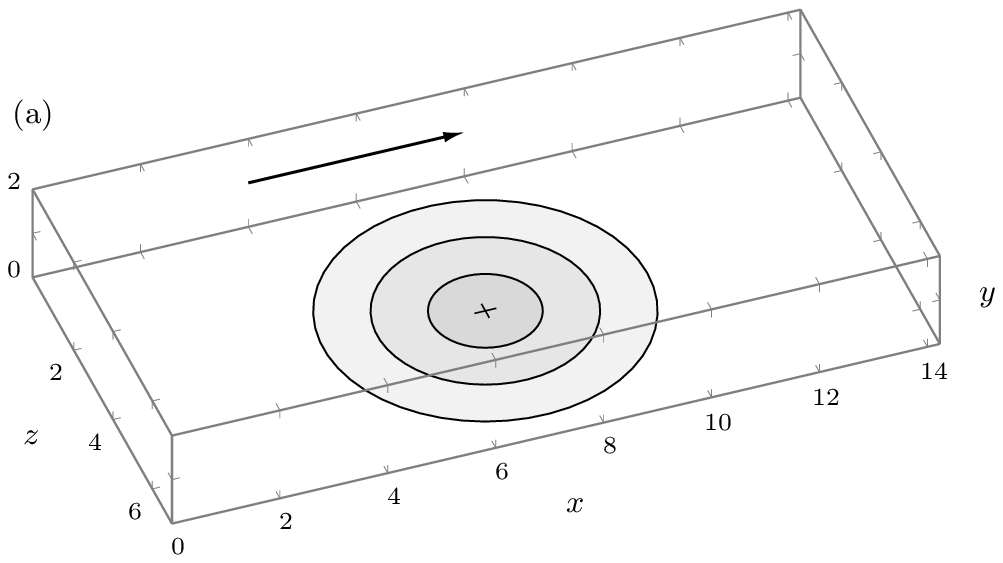}
    \includegraphics[width=\textwidth]{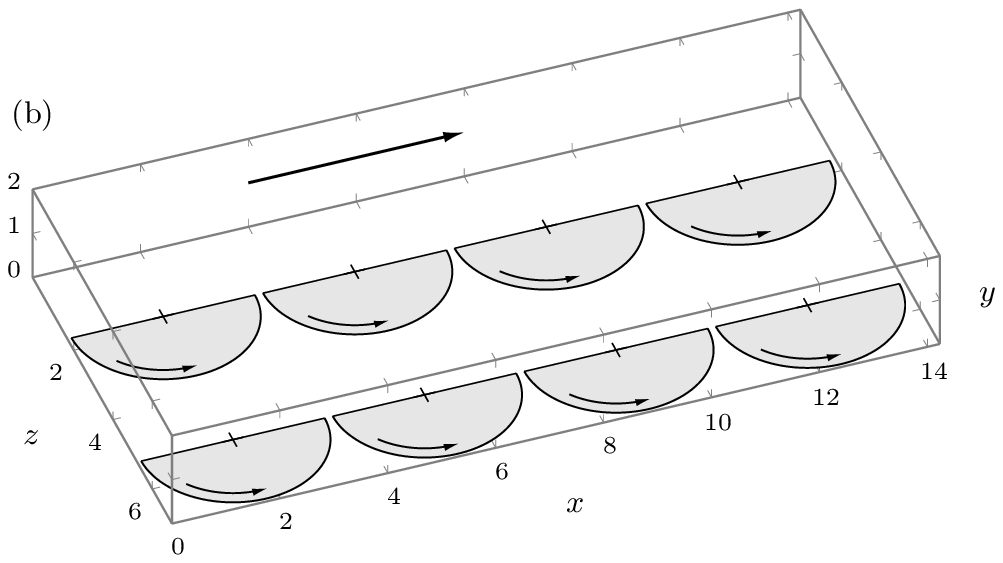}
    \vspace{0.2cm}
\end{subfigure}
\caption{Schematic of (a) the single freely-rotating disc of increasing diameter and (b) the rows and columns of half discs.}
\label{fig:discs}
\end{figure}

Equations \eqref{eq:continuity-cartesian-vector} and \eqref{eq:navier-stokes-cartesian-vector} are subject to the no-slip boundary conditions at the channel walls. The velocity is null over the stationary portions of the walls, while it coincides with the local disc velocity when the fluid passes over a disc. The boundary conditions are time dependent because they are determined by the rigid-body dynamics of the discs, driven by the instantaneous wall-shear stresses exerted by the wall turbulence on the disc surfaces. The dynamics is ruled by a two-way coupling between the fluid and the freely-rotating discs. The Supplementary Material \ref{app:uncoupled-dynamics} discusses an uncoupled model, where the motion of the disc does not feed back to the turbulence dynamics. Two configurations are considered: a full-disc layout where two coaxial discs, one on each wall, are completely exposed to the viscous action of the turbulent flow, and an arrangements of rows and columns of half discs, for which only the right half of the discs is wetted by the fluid. Figure \ref{fig:discs}a shows the configuration of a full disc for three diameters and figure \ref{fig:discs}b shows the configuration of the half discs.

The turbulent channel flow with the freely-rotating discs is studied via direct numerical simulation (DNS). The open-source code Incompact3d \citep{laizet-li-2011,laizet-lamballais-2009} is used to simulate the flows. The code solves the non-dimensional, incompressible NSE using a Chorin-Temam projection method, with time advancement performed using a second-order Adams-Bashforth scheme. The time integration of the equation of motion of the discs is performed using the same numerical scheme of the fluid solver and the same temporal resolution. The spatial discretization employs sixth-order compact finite difference schemes. The Poisson equation for the projection-step pressure is solved in spectral space using fast Fourier transforms on a partially staggered grid. The computational grid is uniformly spaced along the streamwise and spanwise directions and is stretched along the wall-normal direction, ensuring that the near-wall resolution is sufficient to resolve the smallest flow scales accurately. Incompact3d is parallelized using the Message Passing Interface and exhibits excellent scalability, achieved through a 2D domain decomposition \citep{laizet-li-2011}. The code is run on the Cray XC30 ``Archer'' supercomputer of the National Supercomputing Service and the ``ShARC'' cluster at the University of Sheffield. The simulations are run using 1024 parallel computational cores, adopting a $32\times32$ block decomposition. 

The channel flow is assumed to be periodic along the $x$ and $z$ directions and the dimensions of the computational domain in these directions are $L_x$ and $L_z$. For the full-disc cases, the domain dimensions are $L_x=4.53\pi$, $L_y=2$ and $L_z=2.26\pi$ and the number of points defining the numerical grid are $256\PLH 129\PLH 256$ in the $x$, $y$, and $z$ directions, respectively.

All computations are initiated from the laminar Poiseuille channel flow between the two solid stationary walls, perturbed by random noise. The initial flow field evolves to the fully-developed uncontrolled turbulent flow, identified by the main statistics displaying convergence and the total mean stress profile being linear \citep{orlandi-2012}. This fixed-wall fully-developed turbulent flow is used as the initial flow field for the computations of the coupled fluid-disc system. The initial transient flow occurring between the activation of the fluid-disc boundary conditions and the beginning of the fully-developed condition of the flow over freely-moving discs is discarded to acquire meaningful statistics. 

\subsection{Averaging procedures}
\label{sec:averaging}

The time average of a flow variable $f(x,y,z,t)$ over a time interval $[0,T]$ during which $f$ is statistically stationary is defined as:

\begin{equation}
\avertime{f}(x,y,z) = \frac{1}{T}\int^T_0 f(x,y,z,t) \, \mbox{d}t.
\end{equation}

\noindent
For the case where the walls are covered by rows and columns of half discs, the flow is statistically periodic along $x$ and $z$ with period $D_h$, which is the side of the square containing one half disc. A spatial ensemble-averaging operator is therefore defined as:

\begin{equation}
\label{eq:sp-en-avg}
\aversymmetry{f}(\xs,y,\zs,t) 
= 
\frac{1}{N_D}
\sum_{n_x = 1}^{r_D} \sum_{n_z = 1}^{c_D}
f\left(x_s + n_x D_h,y, z_s + n_z D_h,t\right),
\end{equation}
where $0\leq \xs \leq D_h$ and $0\leq \zs \leq D_h$ are the ensemble spatial coordinates, and $r_D$ and $c_D$ are the number of rows and columns of a total number of $N_D = r_D \times c_D$ discs in the computational domain.
Spatial averaging along $\xs$ and $\zs$, i.e., averaging over a square of area $D_h^2$ defining the minimal flow unit containing one disc, is expressed as:

\begin{equation}
\averspace{f}(y,t) = \frac{1}{D_h^2}\int^{D_h/2}_{-D_h/2}\int^{D_h/2}_{-D_h/2} 
\aversymmetry{f}(\xs,y,\zs,t) \, \mbox{d}\xs\mbox{d}\zs.
\end{equation}
The statistical sample is doubled by averaging a quantity across the two channel halves. A capital letter indicates a global average, $F(y) = \averglobal{f}= \averspace{\avertime{f}}$. For example, the global-averaged streamwise velocity is $U(y)=\averglobal{u}$. The velocity field is decomposed as follows, 

\begin{align}
&\textbf{u}(x,y,z,t) = 
\textbf{U}(y) +
\textbf{u}_\textbf{\mbox{d}}(x - \lfloor x/D_h \rfloor D_h, y, z - \lfloor z/D_h \rfloor D_h) + 
\textbf{u}_\textbf{\mbox{t}}(x,y,z,t) 
\label{eq:decomp1},  \\
&
\textbf{u}_\textbf{\mbox{d}} 
= \{u_d, v_d, w_d\} 
= \aversymmetry{\avertime{\textbf{u}}} - \textbf{U}(y)
\label{eq:decomp_ud}, 
\end{align}
where $\textbf{U}(y)=\{U(y),0,0\}$. 

The separation of scales typical of wall-bounded turbulence is measured by the friction Reynolds number $Re_\tau = h^* / \delta^*_\nu=180$, where $\delta^*_\nu = \nu^*/u^*_\tau$ is the near-wall viscous length scale, $u^*_\tau = \sqrt{\averglobal{\tau_{w,x}^*} / \rho^*}$ is the wall-friction velocity, and $\averglobal{\tau_{w,x}^*}=\nu^*\rho^*\text{d}U^*/\text{d}y^*|_{y=0}$ is the global-averaged wall-shear stress in the fixed-wall configuration. Quantities scaled in viscous units, that is, using the fixed-wall $u_\tau^*$ and $h^*$, are denoted by a $+$ superscript. 
The skin-friction coefficient is $C_f=2\averglobal{\tau_{w,x}}/U_b^2$, where $U_b =2/3$. The drag reduction $\mathcal{R}$ is defined as the percentage decrease of the skin-friction coefficient:
\begin{equation}
\mathcal{R}(\%) = 100(\%) \cdot \frac{C_{f,un} - C_f}{C_{f,un}},
\label{eq:drdef}
\end{equation}
where $C_{f,un}$ is the skin-friction coefficient of the uncontrolled flow. We also define a spatially dependent turbulent drag reduction as follows:
\begin{equation}
\mathcal{R}_{xz}\left(x_s,z_s\right)(\%) 
= 
100(\%) \cdot \frac{C_{f,un} - c_f\left(x_s,z_s\right)}{C_{f,un}},
\label{eq:drxz}
\end{equation}
where $c_f\left(x_s,z_s\right) = \left.2 Re_p^{-1} U_b^{-2}\p (U + u_d)/\p y\right|_{y=0}$.
The disc drag-reduction margin $\mathcal{R}_d$ is defined as the reduction of wall friction computed by only taking into account the disc surface.

\section{Modelling of the disc dynamics}
\label{sec:model}

This section presents the modelling of the dynamics of the freely-rotating discs. Unless stated otherwise, this model is valid for both configurations of discs, shown in figure \ref{fig:discs}. In this section, the modelling of the two-way coupled dynamics is presented, while the modelling of the uncoupled dynamics is discussed in the Supplementary Material \ref{app:model-uncoupled-dynamics}.
\begin{figure}
\centering
\includegraphics[width=0.85\textwidth]{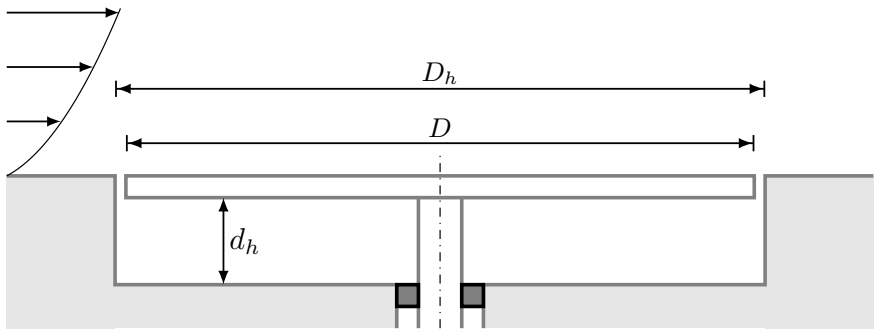}
\vspace{0.5cm}
\caption{Schematic of the disc and its housing. The axis of rotation of the disc is indicated by a dash-dotted line and the dark grey squares represent the ball bearing.}
\label{fig:housing}
\end{figure}
The discs are flush mounted on the two surfaces of the channel, have diameter $D^*$, and are rigid, homogeneous, and free to rotate about their central axis. A schematic of the disc housing is shown in figure \ref{fig:housing}. The discs are characterized by their thickness $b^*$ and material density $\rho_d^*$. The moment of inertia of a disc about its central axis is $I^*= \pi b^* D^{*4} \rho_d^* / 32 $. A thin gap of width $0.05D^*$ is modelled around the edge of each disc to simulate the clearance region between the disc and the stationary channel walls. Within the annular gap the velocity is modelled to decay linearly from its maximum value on the disc edge to zero on the fixed portion of the wall. This treatment of the gap closely represents a real experimental set-up where such gaps would always be present, and it drastically reduces the numerical oscillations that would arise from a discontinuity in the velocity field \citep{ricco-hahn-2013}.

Each disc is fitted with a thin shaft and housed in a cavity in the wall that is deep enough for the upper surface of the disc to be flush with the channel wall. The cavity has thickness $d^*_h$ and is filled with the same fluid of the turbulent channel flow. The shaft is supported by a bearing mounted into a purpose-built hole at the bottom of the disc housing. The disc housing has a similar design to those realized in the experimental study of KK13 and KK14, and in a preliminary experimental apparatus built at the Deutsches Zentrum f{\"u}r Luft- und Raumfahrt (DLR) in G{\"o}ttingen to reproduce the spinning-disc system studied numerically by \cite{ricco-hahn-2013} (M. Rutte and U.G. Becker, private communication). 

It is important to convert the scaled quantities characterizing the turbulent flow and the disc system to physical quantities because one must simulate a flow system that can be realized in a laboratory. One risk when working solely with scaled variables is, for example, to compute a turbulent flow over unrealistically light discs. The very low inertia would lead to exceedingly large angular velocities and thus drag-reduction margins that could not be measured experimentally. Table \ref{tab:waterchan} presents the characteristics of a turbulent channel flow with discs, where the working fluid is water at 20$^\circ$C. Water is preferred to air because the larger dynamic viscosity of water generates a larger frictional torque on the disc surface. The discs must be designed as thin and light as possible in order to minimize their inertia and thus maximize their angular acceleration for a given fluid torque. The disc thickness is chosen to be 0.5mm and we assume the material of the discs to be aluminium. The density of aluminum $\rho_d^*$ is almost three times the density of water, i.e., $\rho_d=2.7$. If air were chosen, the scaled density of the disc would be $\rho_d=2200$. 

Three cavity depths are tested, $d_h=0.1,1$, and 10. The smallest value is chosen to simulate more intense frictional effects caused by the flow in the cavity, while the largest value is selected to simulate a housing torque that is not influenced by the confinement of the cavity. 

\begin{table}
\centering
\caption[width=\linewidth]{Physical quantities of the turbulent water channel flow and the discs.}
\label{tab:waterchan}
\begin{tabular}{@{}c|c|c|c|c|c|c@{}}
\hline
$h^*$ (cm)  & $U^*_b$ (cm/s) & $\nu^*$  (m$^{2}$/s) & $\rho^*$ (kg/m$^3$)           & $Re_p$   & $b^*$ (mm)    & $\rho_d^*$ (kg/m$^3$)   \\ 
\hline
$10$ & $2.8$ 	& $10^{-6}$  	& $1000$  & 4200     & $0.5$  & 
$2700$  \\
\hline
\end{tabular}
\end{table}

\subsection{Dynamics of the rotating discs}
\label{sec:dynamics}

\begin{figure}
\centering
\includegraphics[width=1\textwidth]{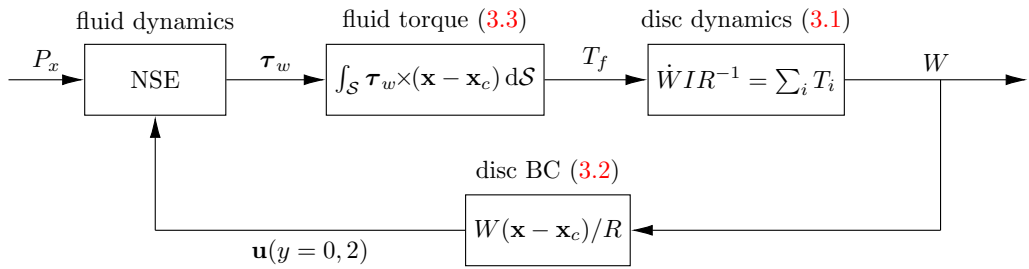}
\vspace{0.1cm}
\caption{Block-diagram of the two-way coupled disc-fluid system as a feedback loop. In this diagram and in figures \ref{fig:uncoupled} and \ref{fig:loop_spec}, the dot indicates the derivative with respect of time and BC stands for boundary conditions.}
\label{fig:coupled}
\end{figure}

The disc motion is described by its angular velocity $\Omega(t)$ or by its disc-tip velocity $W(t) = R \Omega(t)$, where $R$ is the radius of the disc. The disc-tip velocity evolves according to the following dynamical equation:
\begin{equation}
\frac{2I}{D} \frac{\mathrm{d}W}{\mathrm{d}t} = T_f - T_h - T_b, 
\label{eq:sys1}
\end{equation}
where $W$ is positive anti-clockwise, $T_f$ is the torque exerted by the wall-shear stresses of the channel flow on the flow-facing surface of the disc, $T_h$ is the torque arising from the fluid contained in the housing beneath the disc, and $T_b$ is the frictional torque given by the ball bearing. Equation \eqref{eq:sys1} is nonlinear because the torques $T_f$ and $T_h$ depend on $W$. 
The equation of the disc dynamics \eqref{eq:sys1} is solved together with the continuity equation \eqref{eq:continuity-cartesian-vector} and the NSE  \eqref{eq:navier-stokes-cartesian-vector}. The equations are mathematically coupled for two reasons. First, the torque $T_f$, exerted by the wall-shear stresses of the wall turbulence, drives the motion of the disc and is therefore an input into the disc dynamics. Second, the disc motion fixes the boundary conditions of the fluid velocity on the disc surface.
On a disc centred at $(x_c,z_c)$ and rotating with a disc-tip velocity $W$, the velocity boundary conditions for the streamwise and the spanwise velocity components read:
\begin{equation}
 \label{eq:disc_bc}
u(x,y=0,2,z,t) = W(t) (z-z_c) / R, \quad w(x,y=0,2,z,t) = -W(t) (x-x_c) / R,
\end{equation}
for $x,y$ belonging to the surface of the disc, i.e., $(x-x_c)^2+(z-z_c)^2 \leq R^2$.
The coupled system is visualized in figure \ref{fig:coupled} using a block diagram \citep{astroem-murray-2008}. The diagram features the disc-tip velocity as the main output and displays a closed-loop structure due to the two coupling elements, i.e., the fluid torque and the fluid velocity boundary conditions.

For the full-disc cases, only the disc diameter is varied, while the disc thickness, material density, and cavity depth are kept constant. For the case with $D=3.38$, the flow over the isolated full discs, shown in figure \ref{fig:discs}, was compared to the flow over rows and columns of discs with the same diameter and no differences in the statistics were found. 

As in the wind-tunnel study of KK13 and KK14, we consider freely-rotating discs that only have the right spanwise half exposed to the wall turbulence, while the other half is covered by a solid surface and is not exposed to the flow.
The half discs rotate with a finite mean velocity, caused by the averaged fluid torque induced by the mean wall-shear stress on the exposed portion of disc surface. Since the steadily moving discs produce a streamwise slip velocity at the wall, a local flattening of the velocity profile and a reduction of the mean velocity gradient at the wall are expected.
The half-disc simulations, summarized in table \ref{tab:half_cases}, differ in three ways from the experiments of KK13 and KK14. In the experiments, an open free-stream boundary layer is studied, the Reynolds number is larger than in our channel-flow simulations, and the disc diameters are larger in wall units.
The half discs are modelled by applying the fixed-wall condition to the left-hand half of the disc. The disc model is modified by only calculating $T_f$ on the existing disc half and by multiplying $T_h$ by a factor of $1.5$ to account for the friction under the hypothetical plate that would cover the sheltered half of the disc. The latter change implies that the flow under the plate covering the disc half is assumed to be the same as that found in the housing cavity. 

\begin{table}
\centering
\caption{Numerical parameters of the half-disc flow simulations. The numbers of the grid size are the number of grid points along the $x$, $y$, and $z$ directions. The models for $T_h$ are presented in the Supplementary Material \ref{app:torques}.}
\label{tab:half_cases}
\begin{tabular}{@{}lllllll@{}}
\hline
{Case}	& $L_x$ 	& $L_z$ 	& $D^+$  	  & $N_D$ 	& $d_h$ & $T_h$ model	
\\ 
\hline
{HD1}	& $4.53\pi$ & $2.26\pi$ & $605$  	  & 8 & 0.1 & Stewartson	\\
{HD3}	& $4.53\pi$ & $2.26\pi$ & $605$  	  & 8 & 1 	& Stewartson	\\
{HD4}	& $4.53\pi$ & $2.26\pi$ & $605$  	  & 8 & 10 	& Stewartson	\\
{HD2}	& $6.7\pi$  & $3.33\pi$ & $1210$ 	  & 2 & 0.1 & Stewartson	\\
{HD5}	& $4.53\pi$ & $2.26\pi$ & $605$  	  & 8 & 10  & von-K{\'a}rm{\'a}n		\\
\hline
\end{tabular}
\end{table}

\subsection{Modelling of the torques}
\label{sec:torques}

The fluid torque $T_f$ results from the wall-shear stresses exerted by the turbulent flow over the surface $\mathcal{S}$ of a disc and is expressed by the integral:
\begin{equation}
T_f(t) 
= \int_\mathcal{S} \boldsymbol{\tau_w} \times (\mathbf{x}-\mathbf{x}_c) \, \mbox{d}\mathcal{S} 
= \int^{R}_{-R}\int^{\sqrt{R^{2}-(x-x_c)^{2}}}_{-\sqrt{R^{2}-(x-x_c)^{2}}} 
\left[ (z-z_c)\tau_{w,x} - (x-x_c)\tau_{w,z} \right] \, \mbox{d}z \mbox{d}x,
\label{eq:t_f}
\end{equation}
where the streamwise and spanwise components of the wall-shear stress $\boldsymbol{\tau_w}$ are
\begin{align}
	\tau_{w,x} = - \frac{1}{Re_p} \left.\frac{\partial u}{\partial y} \right|_{y=0}, \quad
	\tau_{w,z} =   \frac{1}{Re_p} \left.\frac{\partial w}{\partial y}\right|_{y=0}.
\label{eq:t_f_contrib}
\end{align}
The integral in \eqref{eq:t_f} is evaluated at each time-step and the boundary condition on the disc are updated accordingly, realizing the fully coupled disc-fluid dynamics. The ball-bearing torque is modelled as $T_b = \vert T_b \vert {\rm sgn}(W)$, where $\vert T_b \vert$ is obtained from the specifications of an industrial bearing. The torque exerted by the cavity flow is modelled as $T_h=T_h(Re_p,d_h,W,D)$, using known solutions of flows over spinning discs. Details of the modelling of the torques acting below the disc surface are presented in the Supplementary Material \ref{app:torques}.

\section{Results for the full discs}
\label{sec:results-full}

This section presents the results of turbulent flows over the isolated full freely-rotating discs, depicted in figure \ref{fig:discs}a. The uncoupled-dynamics results are found in the Supplementary Materials \ref{app:uncoupled-physical} and \ref{app:uncoupled-frequency}. For the full-disc simulations, the linear housing torque formula \eqref{eq:th_stewartson} is used. The equation of motion for the disc simplifies to:
\begin{equation}
\frac{\mathrm{d}W}{\mathrm{d}t} = 
\frac{16}{\pi b D^3 \rho_d} \left[T_f(t) - T_b \right] - \frac{W(t)}{Re_p d_h b \rho_d}.
\label{eq:dyn_lin}
\end{equation}

The wall turbulence exerts a finite instantaneous torque on the full disc, which results in an unsteady motion of the disc characterized by its disc-tip velocity $W$. 
The root-mean-square of the disc-tip velocities ($W_{\mbox{rms}}$) and the chi-squared confidence intervals are presented in table \ref{tab:rms_Ds} for the eleven disc diameters studied, ranging between $D^+$=50 and $D^+$=1200. (The subscript rms henceforth indicates the root-mean-square of a quantity). The time average $\avertime{W}$ is null in all the cases because the wall-shear stress has no preferential direction over the disc surface. Small values of the disc-tip velocities are found, the $W_{\mbox{rms}}$ never exceeding the wall-friction velocity. 

\begin{table}
\centering
\caption{$W_{\mbox{rms}}$ for the full-disc coupled cases. The confidence intervals are at the $95\%$ level.}
\begin{tabular}{@{}cccccc@{}}
\hline
$D$  & $D^+$ & $I\times10^3$ & $W_{\mbox{rms}} \times 10^2$  & $W_{\mbox{rms}}^+$  & 	chi-squared confidence on $W_{\mbox{rms}} \times 10^2$  \\ 
\hline
0.28 & 50    & 0.01          & 1.94                   & 0.45        &   $+1.57$, $-1.16$ \\
0.56 & 100   & 0.1           & 2.31                   & 0.54        &   $+1.87$, $-1.39$ \\
0.84 & 150   & 0.7           & 2.01                   & 0.47        &   $+1.63$, $-1.21$ \\
1.13 & 200   & 2.10          & 1.83                   & 0.43        &   $+1.48$, $-1.10$ \\
1.69 & 300   & 10.8          & 1.51	                  & 0.35        &   $+1.22$, $-0.91$ \\
2.25 & 400   & 34.2          & 1.23                   & 0.29        &   $+1.00$, $-0.74$ \\
3.38 & 600   & 173           & 0.95                   & 0.22        &   $+0.77$, $-0.57$ \\
4.44 & 800   & 515           & 0.79                   & 0.19        &   $+0.64$, $-0.48$ \\
5.00 & 900   & 828           & 0.70                   & 0.16        &   $+0.56$, $-0.42$ \\
5.60 & 1000  & 1350          & 0.60                   & 0.14        &   $+0.49$, $-0.36$ \\
6.70 & 1200  & 2670          & 0.58                   & 0.13        &   $+0.47$, $-0.35$ \\
\hline
\label{tab:rms_Ds}
\end{tabular}
\end{table}

\begin{figure}
    \centering
    \begin{subfigure}[h]{\textwidth}
    \centering
    \includegraphics[width=1\textwidth]{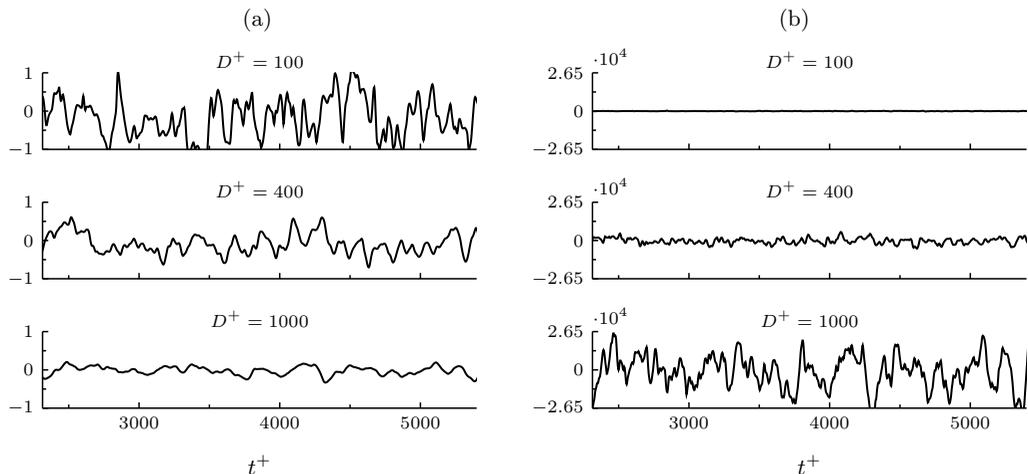}
    \end{subfigure}
    \caption[Time evolution of the torque and disc velocity for three discs of increasing diameter.]{Time evolutions of (a) the disc-tip velocity $W^+$ and (b) the fluid torque $T^+_f$ for discs with three diameters.}
    \label{fig:disc_time_series}
\end{figure}

Time histories of $W$ are shown in figure \ref{fig:disc_time_series}a for three diameters. The discs oscillate randomly around $W=0$ and the variance of $W$ is larger at small diameters, the disc oscillations becoming less intense as the diameter increases. The time histories of the torque in the same time interval, shown in figure \ref{fig:disc_time_series}b, behave in the opposite way, the variance of $T_f$ growing rapidly with the diameter. 

Because of the small angular displacements, no drag reduction occurs. \citet{ricco-quadrio-2008} for spanwise-wall oscillations, \citet{quadrio-ricco-2011} for wall-travelling waves and \citet{ricco-hahn-2013} for rotating discs demonstrated that a minimal wall velocity, about $W^+=1$, is required for the wall turbulence to be affected by the wall motion and to experience drag reduction. If the wall displacement is too small, the viscous effects are confined in a very thin layer because the wall-normal momentum diffusion from the wall is limited. In the freely-rotating disc case, $W_{\mbox{rms}}$ is always smaller than the minimal wall velocity and is thus expected that the skin-friction drag is unaffected.

\begin{figure}
\centering
\begin{subfigure}[l]{\textwidth}
    \includegraphics[width=1\textwidth]{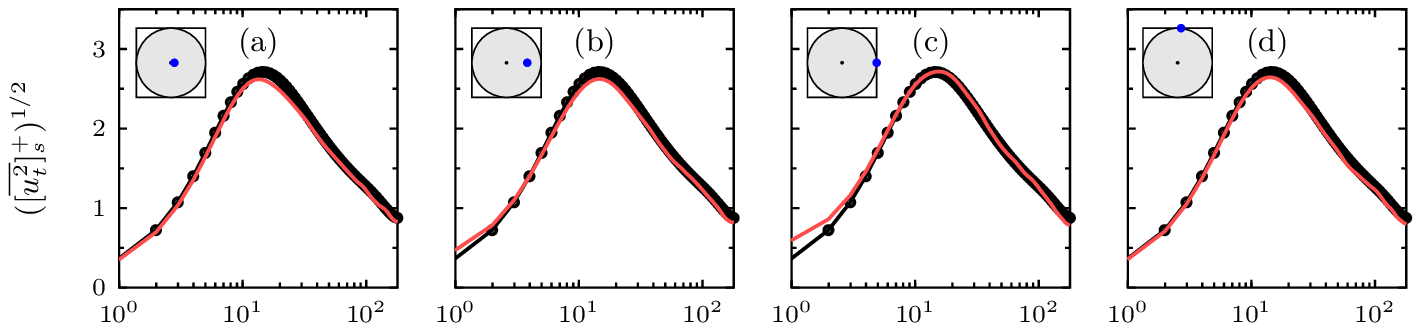}
\end{subfigure}
\begin{subfigure}[l]{\textwidth}
    \includegraphics[width=1\textwidth]{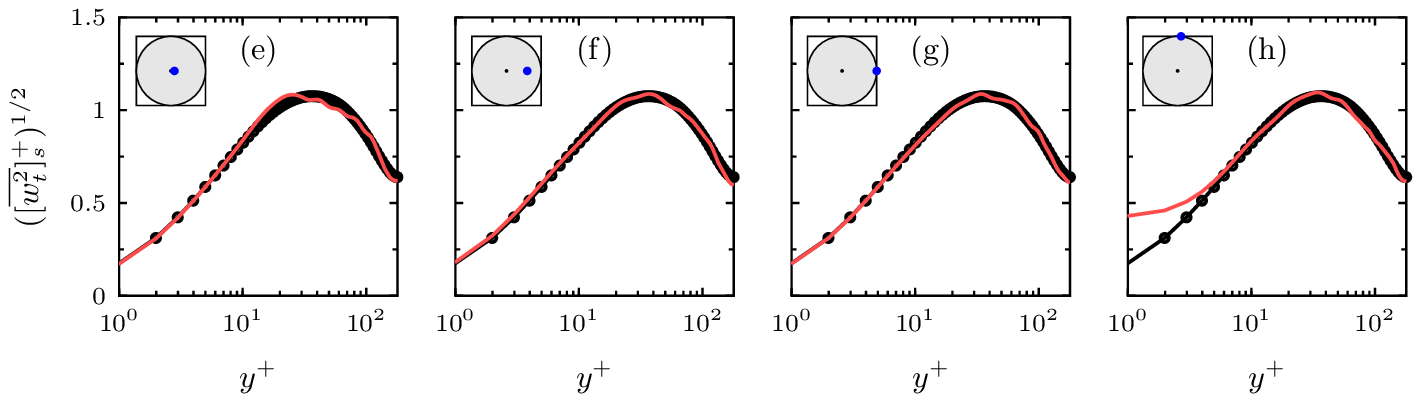}
\end{subfigure}
\caption{
Wall-normal profiles of the root-mean-square of the streamwise (a--d) and spanwise (e--h) velocity fluctuations at different positions on the freely oscillating disc for the case with $D=0.84$ (\raisebox{1.5pt}{\includegraphics[height=.1\baselineskip]{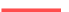}}) and the reference channel-flow case (\includegraphics[height=.5\baselineskip]{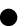}). The locations are shown on the top left of each graph. The variables are scaled with the wall units of the reference channel flow.
}
\label{fig:stat-pos-free}
\end{figure}

The profiles of the root-mean-squares of the streamwise and spanwise velocity fluctuations are reported in figure \ref{fig:stat-pos-free}. The wall-normal extension of the oscillation-induced flow is confined to the viscous sublayer and cannot modify the structure of near-wall turbulence for $y^+\ge 5$. The mean velocity, turbulent Reynolds stresses and root-mean-square  of the wall-normal velocity fluctuations (not depicted in figure \ref{fig:stat-pos-free}) are indistinguishable from those of the reference channel flow.

\begin{figure}
    \centering
    \hspace{-.06\textwidth}
    \begin{subfigure}[h]{\textwidth}
    \centering
    \includegraphics[width=0.95\textwidth]{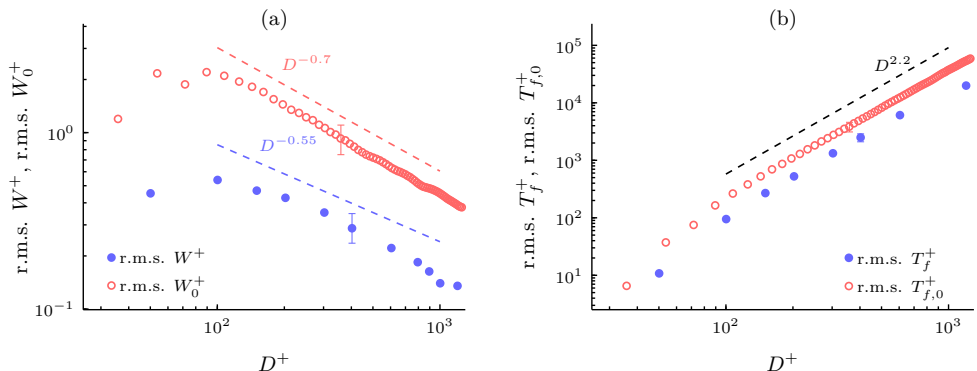}
    \end{subfigure}
    \caption[Dependence of $W^+_{\mbox{rms}}$ and $T^+_{f,\mbox{rms}}$ on the disc diameter.]{
    Root-mean-square values of (a) the disc-tip velocities $W^+$ in the coupled case and $W^+_0$ in the uncoupled case, and (b) the fluid torque $T^+_f$ as functions of the disc diameter. The time-averaging interval is $2500h^*/U^*_p$ or $19300 \nu^*/u_\tau^{*2}$. The blue full circles denote the coupled-dynamics data and the red empty circles symbols denote the uncoupled-dynamics data, discussed in the Supplementary Material \ref{app:uncoupled-dynamics}.
    }
    \label{fig:rms_Ds}
\end{figure}

Figure \ref{fig:rms_Ds} shows the values of $W_{\mbox{rms}}$ and $T_{f,\mbox{rms}}$. The blue symbols denote the coupled-dynamics results, summarized in table \ref{tab:rms_Ds}, and the red symbols denote the uncoupled-dynamics results, discussed in the Supplementary Material \ref{app:uncoupled-dynamics}.
For $D^+$\textless100, $W_{\mbox{rms}}$ grows with the disc diameter, while $W_{\mbox{rms}}$ decreases as $D^{-0.55}$ for 100\textless$D^+$\textless900. The maximum $W_{\mbox{rms}}$ at $D^+$=100 occurs because, for very small $D$, the torque is too small to exert the required power to move the discs, while for large discs, the inertia of the disc thwarts the action of the torque. The optimum diameter is comparable with the characteristic spanwise spacing of the low-speed streaks \citep{kline-etal-1967}. The torque $T_{f,\mbox{rms}}$ is instead a monotonically increasing function of the diameter. For 0\textless$D^+$\textless100, $T_{f,\mbox{rms}}$ grows more rapidly with $D$ than for larger diameters and is proportional to $D^{2.2}$ for 100\textless$D^+$\textless900. The uncoupled-dynamics data reveal that the two-way fluid-disc coupling has an attenuating effect on the kinetic energy of the disc.

\begin{figure}
\centering
    \includegraphics[width=0.49\textwidth]{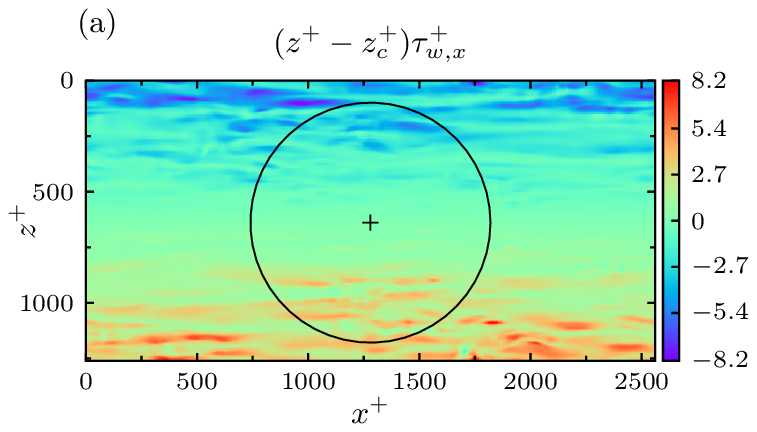}
    \includegraphics[width=0.49\textwidth]{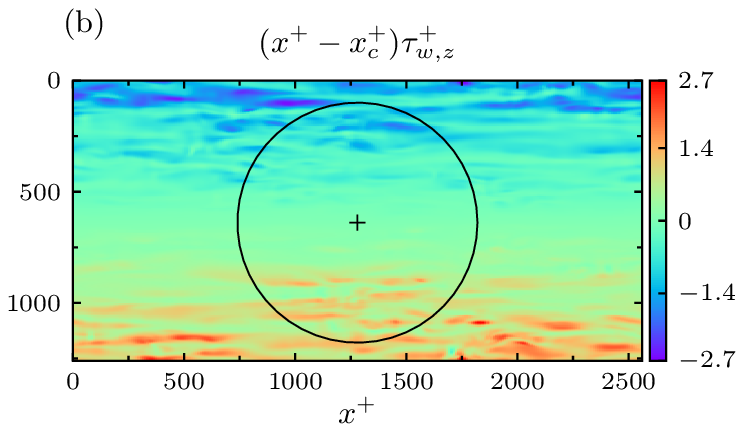}
\caption{Instantaneous snapshots of (a) the torque contribution given by the streamwise wall-shear stress and (b) the torque contribution given by the spanwise wall-shear stress, on a disc with $D=6$.}
\label{fig:torque}
\end{figure}

\begin{figure}
    \centering
    \begin{subfigure}[h]{.49\textwidth}
    \centering
        \includegraphics[width=1\textwidth]{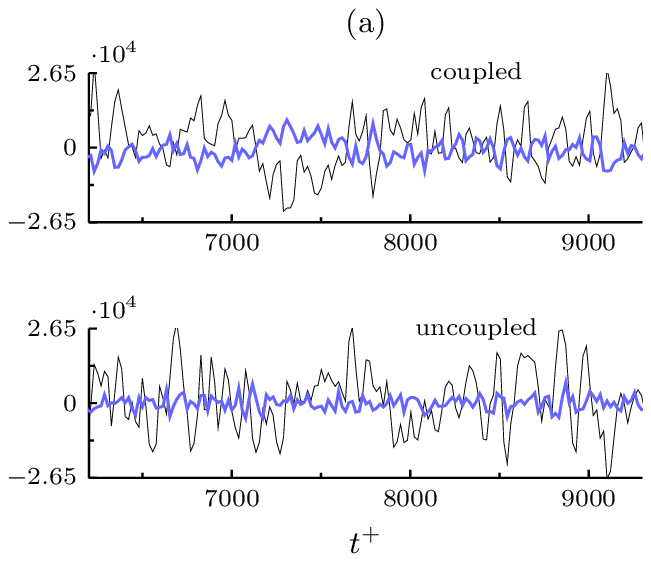}
        \vspace{0.2cm}
    \end{subfigure}
    \begin{subfigure}[h]{.49\textwidth}
    \centering
        \includegraphics[width=1\textwidth]{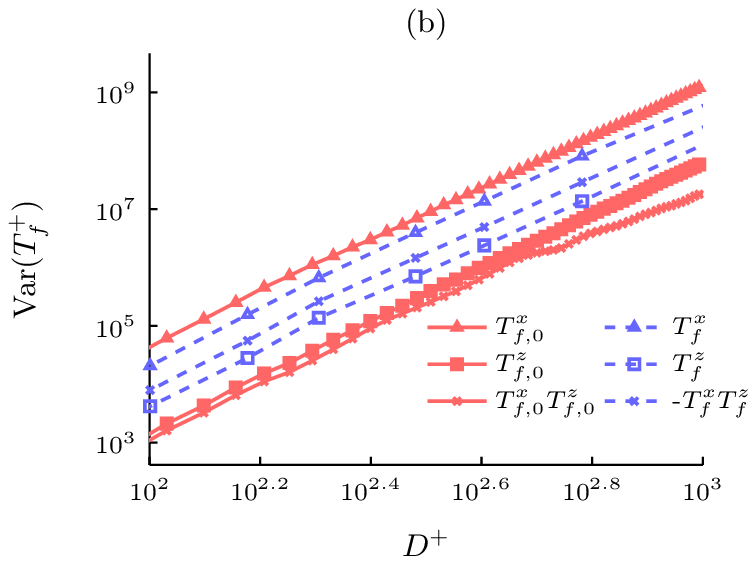}
    \end{subfigure}
    \caption{
    Spanwise and streamwise components of the fluid torque.
    (a) Time-series of $T^{x+}_f(t)$ (black) and $T^{z+}_f(t)$ (colour) for the uncoupled (top) and the coupled case (bottom). (b) Mean-square values and covariance as a function of the disc diameter.
    The blue series indicate the data from the coupled simulations and the red series indicate the data are from the uncoupled simulations, discussed in the Supplementary Material \ref{app:uncoupled-dynamics}.
    }
    \label{fig:tf_components}
\end{figure}

The torque can be written as $T_f = T^x_f + T^z_f$, where $T_f^x$ and $T_f^z$ are given by the streamwise and spanwise wall-shear stresses, $\tau_{w,x}$ and $\tau_{w,z}$, respectively. Figure \ref{fig:torque} visualizes instantaneous snapshots of the torque contributions. The component $T_f^x$ is much larger than $T_f^z$ and most intense at the disc sides.
Figure \ref{fig:tf_components}a (top) shows that $T^x_f$ and $T^z_f$ have opposite sign for most of their time histories. The torque variance is expressed as:
\begin{equation}
\mbox{Var}(T_f) = \avertime{{T_f^x}^2} + \avertime{{T_f^z}^2} + 2 \avertime{T_f^x T_f^z},
\label{eq:variance_decomp}
\end{equation}
where the last term is the covariance of the streamwise and spanwise terms. The dependence of the right-hand side terms of equation \eqref{eq:variance_decomp} on the disc diameter is shown in figure \ref{fig:tf_components}b by the open blue symbols. All three components share a very similar growth with the disc diameter. The streamwise components dominate over the other two components, contributing to more than 90\% of the total variance. The covariance is negative and larger in absolute value than the spanwise component.

\begin{figure}
    \centering
    \begin{subfigure}[h]{\textwidth}
    \centering
    \includegraphics[width=1\textwidth]{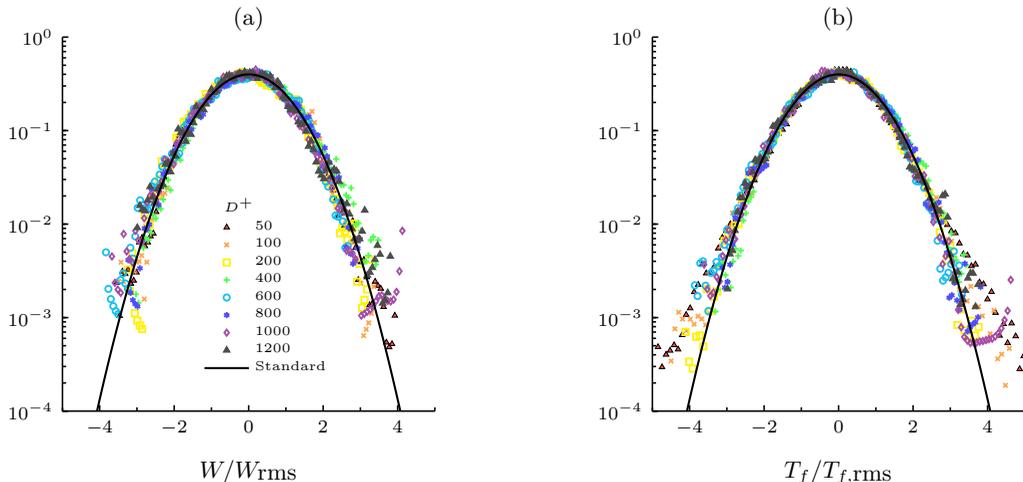}
    \end{subfigure}
    \caption{
    Standardized histogram plots of (a) $W$ and (b) $T_f$ for coupled-dynamics simulations. The solid line denotes the standard normal distribution.
    }
    \label{fig:pdf_Ds}
\end{figure}

The standardized histograms plots for $W$ and $T_f$ are shown in figure \ref{fig:pdf_Ds} and compared to the standard normal distribution. The disc-tip velocity $W$ fits the normal distribution well for all the diameters. The $T_f$ distribution is instead wider than the normal distribution at the tails for the small diameters. The graph of $T_f$ for $D^+=50$ has heavy tails, with an estimated value of the standardized fourth moment larger than four. This disc size  is smaller than the spanwise integral length scale of the wall-shear stress, about $100 \delta_\nu^*$, and thus a possible explanation for the tail behaviour is that the torque fluctuations depend on localized instantaneous extreme events, impacting on the positive tails of the wall shear-stress distribution \citep{hu-etal-2006}. For larger discs, the extreme shear-stress events are likely to play a more marginal role because they are averaged over a larger wall region. Although the wall turbulence is non-Gaussian \citep{davidson-2004}, the very good agreement between the $W$ distribution and the normal distribution is likely to be due to the filtering of the turbulence signal given by the disc dynamics equation \eqref{eq:sys1}. The extreme rare bursting and sweeping events that characterize the wall-shear stress and cause the heavy tails of the torque distribution are too short-lived to accelerate the disc so as to generate corresponding extreme values of disc-tip velocity $W$. It would be interesting to verify whether the $W$ distribution deviates from the normal one if turbulent flows at larger Reynolds numbers or lighter discs are considered.

\begin{figure}
    \centering
	\begin{subfigure}[h]{\textwidth}
	\centering
	\includegraphics[width=1\textwidth]{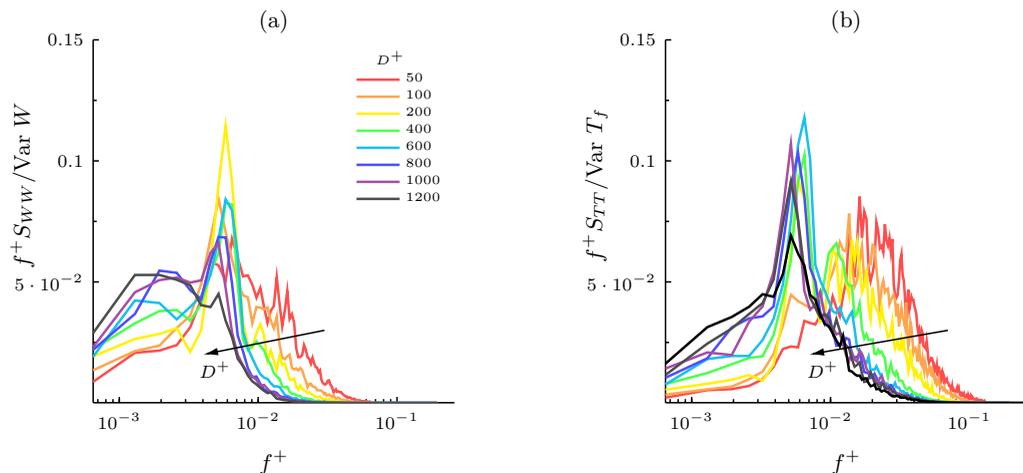}
	\end{subfigure}
    \caption{
    Power spectral densities, denoted by $S$ with subscripts indicating the quantity and shown in premultiplied form. (a) Disc-tip velocity $W$ and (b) fluid torque $T_f$. The plots are normalized by the total power.
    }
    \label{fig:psd_Ds_scaled}
\end{figure}

The Power Spectral Densities (PSD) for the disc-tip velocity and the fluid torque, shown in figure \ref{fig:psd_Ds_scaled}, reveal which temporal scales of the wall turbulence are selected by the disc dynamics. Normalizing the PSD graphs with the total power does not result in a collapse of the curves for all diameters. An excellent overlap for $W$ and $T_f$ is observed for the cases with the four largest diameters, for $D^+=800-1200$, at frequencies that are larger than the one corresponding to the maximum energy. For $D^+=50-400$, most of the energy is contained in the higher frequencies. As the diameter increases, the spectrum shifts to lower frequencies. For these diameters, no distinct peak is found for $W$ in the range $D^+=50-400$ as the energy spreads over a relatively broad high-frequency range.
The same discussion holds for the $T_f$ PSD in the range $f^+=0.01-0.03$, but for $D^+=200$ and $400$ two distinct maxima are present. For moderate diameters, in the range $D^+=600-800$, a narrow peak occurs for $W$ and $T_f$ at around $f^+=0.005$, which corresponds to a period of about $200 \nu^*/u_\tau^{*2}$. As the diameter increases, the spectral $W$ peak disappears and its energy is distributed more evenly at frequencies $f^+$\textless0.004. The torque $T_f$ instead maintains the peak at about $f^+=0.005$.

\section{Results for the half discs}
\label{sec:results-half}

The main difference between the full discs and the half discs is the finite average velocity in the half disc case due to the driving action of the mean turbulent flow. Once the new fully-developed regime is established, the disc inertia can be neglected because the half-disc $\overline{W}$ is determined by the steady-state balance between the fluid torque and the frictional housing torques $\overline{T}_f(\overline{W}) = \overline{T}_h(\overline{W}) + T_b$. In analogy with studies on drag reduction by hydrophobic walls where a local slip velocity is used \citep{min-kim-2004,busse-sandham-2012}, we define the average streamwise slip velocity as $U_{s,d}=\averspace{\avertime{u}}_d$, where the subscript $d$ denotes the spatial average over the moving disc surface only. The slip velocity is related to $\overline{W}$, i.e., $U_{s,d} = 4 \overline{W} / 3 \pi$.
No averaged spanwise slip velocity occurs because the discs are rigid, that is, the left-oriented spanwise velocity distribution downstream of the disc centre is antisymmetric to the right-oriented spanwise velocity distribution upstream of the disc centre. The magnitude of the average disc velocity of each half is equal to half of $U_{s,d}$.

\subsection{Turbulent drag reduction}
\label{sec:drag-reduction}

\subsubsection{Performance and spatial distribution}

As shown in table \ref{tab:half_results}, cases HD1 and HD2 return small drag-reduction margins of around $\mathcal{R}=2\%$, which become almost $\mathcal{R}_d=5\%$ when the half-disc surface is considered. The disc-tip velocities and the drag-reduction margins are the same for these cases, although the disc diameters are different, suggesting a direct link between the slip velocity and the drag reduction, as for flows over hydrophobic surfaces \citep{min-kim-2004}. The best performing cases are HD4 and HD5, with $\mathcal{R}_d=20\%$ and $\mathcal{R}_d=14\%$, respectively.

\begin{table}
\centering
\caption{Results from the half-disc simulations.}
\label{tab:half_results}
\begin{tabular}{@{}llllll@{}}
\hline
{Case} & $D^+$	&	$\overline{W}^+$ 	&	$U^+_{s,d}$ & $\mathcal{R}(\%)$ & $\mathcal{R}_d(\%)$ \\ 
\hline
HD1 		& 605	&	0.98  &  0.42   &  2.1	  & 4.9		\\		
HD2 		& 1210	&	0.98  &  0.42   &  2.1	  & 4.9		\\		
HD3 		& 605	&	1.42  &  0.60   &  2.8	  & 8.6		\\		
HD4 		& 605	&	4.85  &  2.06   &  5.6	  & 19.9	\\		
HD5 		& 605	&	2.49  &  1.05   &  5.1	  & 13.9	\\	
\hline
\end{tabular}
\end{table}

\begin{figure}
\centering
\begin{subfigure}[l]{.6\textwidth}
\centering
	\includegraphics[width=1\textwidth]{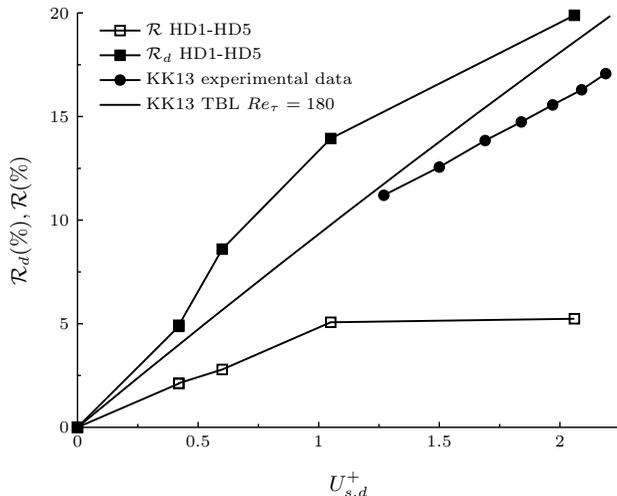}
\end{subfigure}
\caption{
Drag-reduction margins $\mathcal{R}_d$ and $\mathcal{R}$ as functions of the slip velocity $U_{s,d}^+$. The simulations refer to cases HD1-HD5 in table \ref{tab:half_results} and the boundary-layer experimental data are by KK13.
}
\label{fig:lplus_vs_dr}
\end{figure}
\begin{figure}
\centering
\begin{subfigure}[l]{.33\textwidth}
\centering
	\includegraphics[width=1\textwidth]{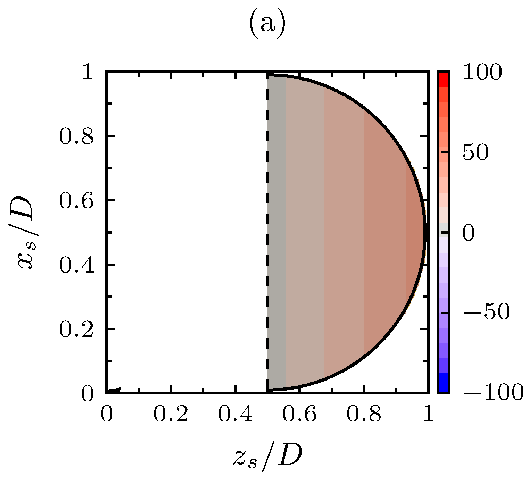}
\end{subfigure}~
\hspace{.2cm}~
\begin{subfigure}[l]{.30\textwidth}
\centering
	\includegraphics[width=1\textwidth]{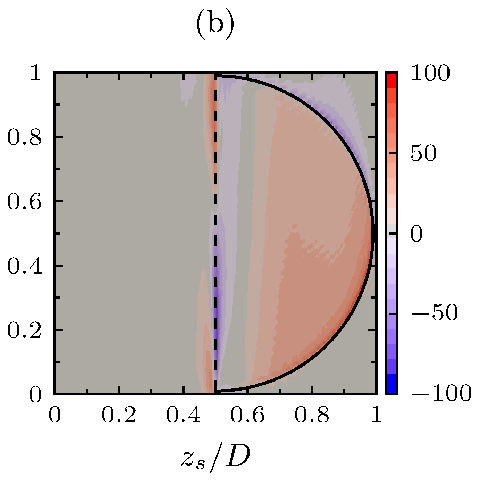}
\end{subfigure}~
\begin{subfigure}[l]{.30\textwidth}
\centering
	\includegraphics[width=1\textwidth]{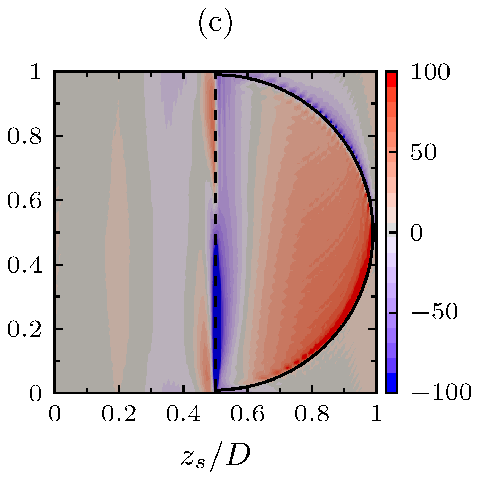}
\end{subfigure}
\caption{
Spatial distribution of the drag-reduction margin $\mathcal{R}_{xz}$ on a single, ensembled-averaged flow unit. (a) Estimation from experimental data by KK13 (fastest rotating case, the region where no data are available is in white), (b) case HD5, (c) case HD4. 
}
\label{fig:haf-dr-xz}
\end{figure}

Figure \ref{fig:lplus_vs_dr} shows the drag-reduction margin as a function of the slip velocity, including the experimental data of KK13. KK13 estimated the drag-reduction level by using an empirical formula based on a correlation of the skin-friction coefficient in the fixed-wall case, as discussed in the Supplementary Material \ref{app:kk}. We use this empirical formula to estimate $\mathcal{R}_d$ as a function of $U_s^+$ by numerically integrating the relation \eqref{eq:kkrxz} for our Reynolds number $Re_\tau=180$. As KK13 did not measure the wall-shear stress on the disc or in its proximity, no experimental data exist for a precise quantitative comparison. Nevertheless, it is useful to compare our numerical results with the predictions given by KK13's formula. In both the experiment and the simulations, $\mathcal{R}_d$ is positively correlated to $U^+_{s,d}$, although the estimated values are lower than those obtained by DNS. These differences may be due to different housing and bearing frictions, Reynolds numbers, and geometry, i.e., a boundary-layer flow in the experiments and a channel flow in our case. Another cause of discrepancy can be the attenuation of the near-wall turbulence intensity caused by the streamwise slip, which is not accounted for by KK13's model.

As $U^+_{s,d}$ increases, $\mathcal{R}$ first increases linearly and then levels off at higher disc velocities. On the stationary wall region (amounting to $63\%$ of the total wall area) neither drag reduction nor drag increase is measured for $U_{s,d}^+\approx1$ (case HD5), while a $3\%$ drag increase is observed for $U_{s,d}^+\approx2$ (case HD4). Although $\mathcal{R}_d$ on the rotating half disc increases by $6\%$ from HD5 to HD4, the non-negligible drag increase on the rest of the wall surface causes $\mathcal{R}$ to only grow by about $0.6\%$.

Figure \ref{fig:haf-dr-xz}a reports $\mathcal{R}_{xz}$, the spatial distribution of the drag-reduction margin,  for the best-performing experiment of KK13, obtained by using the empirical correlation \eqref{eq:kk_shearstress}. The estimated wall-shear-stress reduction is streamwise homogeneous and proportional to the local streamwise velocity, increasing linearly with $z$. The prediction of KK13 thus fails to capture the streamwise dependence because the assumption is that the drag reduction is solely caused by the streamwise wall slip induced by the spinning disc.

The maps of $\mathcal{R}_{xz}$ for the numerical cases HD4 and HD5, shown in the graphs \ref{fig:haf-dr-xz}b--c, reveal a non-homogeneous spatial distribution along both $x$ and $z$. The drag-reduction margin increases moving away from the centreline as predicted by KK13's model because the local streamwise slip velocity grows as the right side of the disc is approached. The maximum wall-shear-stress reduction occurs over the lower part of the rotating disc, while the maximum drag increase occurs in a region that is narrow in the spanwise direction, near the centreline, and upstream of the disc centre. This increase of wall friction might be an effect of the modelled discontinuity at the centreline. As the slip velocity doubles from $U_{s,d}^+=1$ for HD5 to $U_{s,d}^+=2$ for HD4, the drag-increase region on the disc surface and near the centreline becomes wider and more intense, while the drag-reduction margin also grows, especially near the disc tip. The improvement of drag reduction as the disc spins faster is more significant than the increase of drag.  
These two contrasting effects are the reason why the drag-reduction margin over the disc surface does not double as does the disc-tip velocity, as shown in figure \ref{fig:lplus_vs_dr}. In the HD5 case with low $U_{s,d}^+$, the wall-shear stress is almost unchanged over the fixed portion of the surface. In the faster HD4 case, a narrow region of drag reduction is instead observed close to the centreline, while a thin area of increased drag is observed around the circumference and immediately downstream of the disc, where the coupled-dynamics boundary conditions change to fixed-wall no-slip. This increased wall-shear stress is the reason for the enhanced drag over the stationary portion of the wall, which balances the drag reduction over the disc. No drag reduction is found downstream of the disc, as claimed by KK13 and KK14.

\subsubsection{Analogies with other drag-reduction techniques}
\label{sec:analogies}

An interesting case for comparison is the oil-channel experimental campaign of \citet{bechert-etal-1996}. In these experiments, a section of the channel wall was replaced with a tensioned steel belt, free to move in the streamwise direction under the action of the wall-shear stress.  By recording the shear force on the passive wall section and the speed of the belt, \citeauthor{bechert-etal-1996} related the wall slip velocity to the drag reduction. Slip velocities in the range $U_s$$\approx$0.05-0.1 resulted in drag reductions between 5\% and 10\%. Our simulations achieve higher $\mathcal{R}$ for a similar slip velocity (i.e., for case HD5 with $U_{s,d}$=0.05, $\mathcal{R}_d$=14\%, while for case HD4 with $U_{s,d}$=0.08, $\mathcal{R}_d$=20\%). The differences are likely due to the higher Reynolds numbers in the experiments and to the experimental challenges associated with ensuring that the moving belt remained taut and horizontal.

It is also useful to compare the disc results with those produced by hydrophobic surfaces. 
A relationship exists between the hydrophobic-wall $U_s^+$ and the drag-reduction level \citep{rastegari-akhavan-2018}. By using the data of \cite{min-kim-2004}, $U_s^+=0.6$ leads to $\mathcal{R}=7.8\%$ and $U_s^+=1$ gives $\mathcal{R}=12.4\%$. These hydrophobic-wall values pertain to the whole wall, but they must be compared with our $\mathcal{R}_d$ values because in our case the slip is confined to the disc surface. The hydrophobic-wall $\mathcal{R}$ values are in good agreement with our $\mathcal{R}_d$ values in table \ref{tab:half_results}.
\cite{rastegari-akhavan-2018} also used a formula, their equation (5.3), that relates the drag-reduction level to the shift of the quantity $B$, the intercept of logarithmic-law formula for the mean velocity profile when scaled using the drag-reducing friction velocity, due to the change of wall-shear stress. By using the shifts $\Delta B$ for our cases HD4 and HD5 in their formula, we obtain $\mathcal{R}=5.6\%$ for the HD4 case and $\mathcal{R}=5.1\%$ for the HD5 case, in perfect agreement with our numerical values in table \ref{tab:half_results}. 

The spanwise disc velocity varies linearly from its negative minimum at the upstream disc edge to its positive maximum at the downstream disc edge. A similar spanwise-wall velocity distribution was used by \cite{viotti-quadrio-luchini-2009}, where the wall forcing was spanwise only, sinusoidal and distributed over the whole wall. Large drag-reduction margins exceeding 40\% were obtained. It is therefore reasonable to compare the passive disc technique with \cite{viotti-quadrio-luchini-2009}'s case and ask whether the spanwise disc velocity affects the drag-reduction performance and whether it is the reason behind the drag reduction computed via DNS being larger than that obtained by  streamwise-only KK13's model, as shown in figure \ref{fig:lplus_vs_dr}.
To address this question, two further numerical cases are considered by modifying case HD4. In one case, the disc streamwise velocity is assigned the value corresponding to the velocity arising passively in case HD4, while the spanwise velocity is artificially set to zero. In the other case, the spanwise velocity is instead prescribed and the streamwise velocity is null. The results in table \ref{tab:stream-span} show that, in terms of the overall drag-reducing performance, the streamwise-only case leads to larger drag reduction than the HD4 case, both over the disc surface and by averaging over the entire wall. The purely spanwise forcing instead results in very small drag reduction on the disc surface and drag increase over the entire wall. One reason for the small drag reduction in the spanwise-only case is that only part of the wall is forced along the spanwise direction, whereas, in \cite{viotti-quadrio-luchini-2009}'s case, the wall velocity is spanwise homogeneous. Another reason is the low value of spanwise wall velocity, which is not large enough to generate a sufficiently thick spanwise viscous layer, required to alter the wall turbulence effectively and to lead to drag reduction \citep{quadrio-ricco-viotti-2009,quadrio-ricco-2011}. The analysis on the flow statistics of these two cases, found in \S\ref{sec:flow-stats}, further elucidates the dynamics of the spanwise-only case. We conclude that the two forcing effects are not additive, the streamwise disc slip is the main responsible cause for drag reduction, and the spanwise wall velocity only plays a marginal role in the overall drag-reduction dynamics. We note that the role of the spanwise wall velocity in the half-disc technique is thus different from that in hydrophobic surfaces, where a spanwise slip is always detrimental to the overall drag-reduction effect.

\begin{table}
\centering
\caption{Statistics of the single-component simulations and of case HD4.}
\begin{tabular}{@{}lcccc@{}}
\hline
Flow case       & $W^+$   & $U_{s,d}^+$   & $\mathcal{R}_d(\%)$ & $\mathcal{R}(\%)$  \\ 
\hline
streamwise only & 4.9     & 2.1       & 25                  &   7.9              \\
spanwise only   & 4.9     & -         & 0.8                 &   -1.9             \\
HD4             & 4.9     & 2.1       & 20                  &   5.2              \\
\hline
\label{tab:stream-span}
\end{tabular}
\end{table}

\subsubsection{Water or air as working fluid}

As discussed in \S\ref{sec:model} it would be better to use water instead of air as working fluid because of the larger dynamic viscosity of water, which would cause larger wall-shear stresses on the disc surface, slip velocities and drag-reduction margins. Nevertheless, it is of interest to compare the behaviour of the half discs in the two fluids. To conduct this comparison, it must be decided which parameters are kept constant. We assume that the same discs are used (same diameter, thickness, and material), the half-height of the channel $h^*$ is the same, and the flows are at the same $Re_p$.

The ratio of the angular accelerations for the water and air flows can be estimated by using (\ref{eq:sys1}) in dimensional form, neglecting the friction losses, 

\begin{equation}
\label{eq:dwdt}
\frac{\mathrm{d}W^*/\mathrm{d}t^*|_{\mbox{water}}}{\mathrm{d}W^*/\mathrm{d}t^*|_{\mbox{air}}} 
= 
\frac{\nu^*|_{\mbox{water}} \ \mu^*|_{\mbox{water}}}{\nu^*|_{\mbox{air}} \ \mu^*|_{\mbox{air}}}
\approx 3,
\end{equation}
where $\mu^*$ is the dynamic viscosity. 

For the full discs, once the oscillatory regime has established, the different angular accelerations would not impact on the flow dynamics because the typical disc tip velocity would be smaller than the friction velocity in both air or water and drag reduction would not occur in either case. 
The different accelerations would instead impact on the transient dynamics of the half discs from the stationary condition to the drag-reduction regime. The half discs in water would accelerate at about three times the angular velocity of the half discs in air and it would thus take less time in water for the half discs to reach the fully-developed conditions.
Once the fully-developed condition of the half discs is established, the averaged angular acceleration is zero because the fluid torque is balanced by the friction torque below the surface of the disc. By estimating the fluid torque as $T_f^* \sim \tau_w^* R^{*3}$, the friction torque as $T_h^* \sim W^* R^*$, and by extracting the wall-shear stress from the friction Reynolds number, it follows that
\begin{equation}
\label{eq:w}
\frac{W^*|_{\mbox{water}}}{W^*|_{\mbox{air}}} 
= 
\frac{\nu^*|_{\mbox{water}} \ \mu^*|_{\mbox{water}}}{\nu^*|_{\mbox{air}} \ \mu^*|_{\mbox{air}}}
\approx 3,
\end{equation}
in agreement with equation \eqref{eq:dwdt} for the acceleration. In viscous units, the ratio becomes 
\begin{equation}
\label{eq:w+}
\frac{W^+|_{\mbox{water}}}{W^+|_{\mbox{air}}} 
= 
\frac{\mu^*|_{\mbox{water}}}{\mu^*|_{\mbox{air}}}
\approx 50.
\end{equation}
It follows from figure \ref{fig:lplus_vs_dr} that, for the same conditions discussed earlier, the drag reduction in air would be almost negligible compared with the drag reduction in water.

\subsection{Flow visualizations}
\label{sec:flow-vis}

\begin{figure}
\centering
\hspace{-1cm}~
\begin{subfigure}{.5\textwidth}
\centering
	\includegraphics[width=1\textwidth]{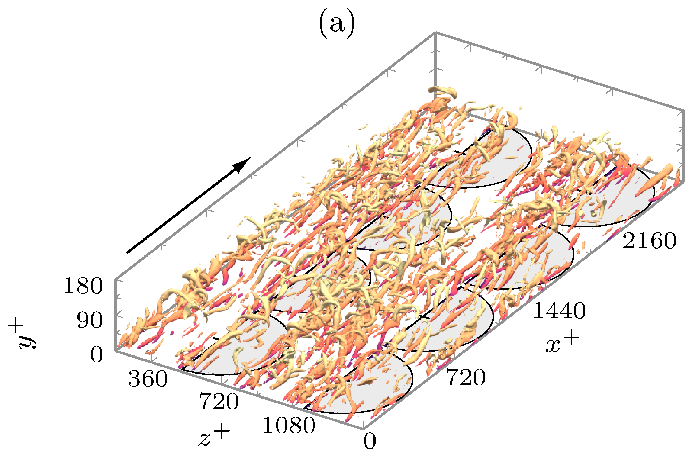}
\end{subfigure}~
\begin{subfigure}{.5\textwidth}
\centering
	\includegraphics[width=1\textwidth]{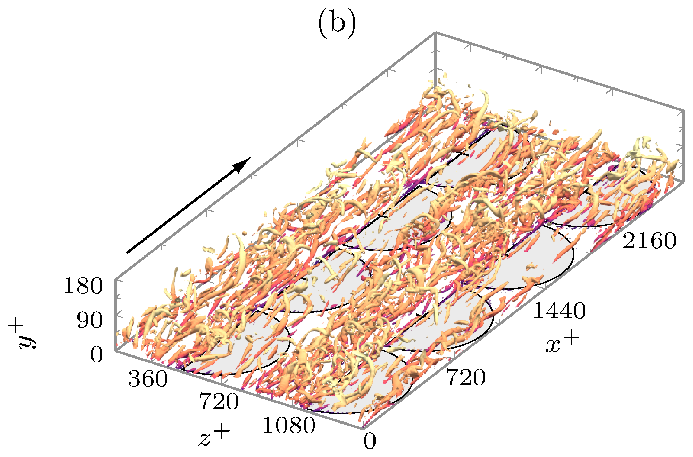}
\end{subfigure}
\caption{
Instantaneous iso-surfaces of the invariant $\lambda_2^+=-0.009$ \citep{jeong-hussain-1995}, coloured by the streamwise velocity component. (a) case HD5, (b) case HD4.
}
\label{fig:lambda2_hd}
\end{figure}

Figure \ref{fig:lambda2_hd} shows instantaneous flow visualization of the fine-scale vortical structures using the $\lambda_2$ invariant method \citep{jeong-hussain-1995}. A persistent reduction in the size and number of the vortices occurs along the streamwise direction above the spinning discs, in contrast with the stationary-wall part that is visually indistinguishable from the reference fixed-wall flow. The effect is more pronounced for case HD4 as the slip velocity $U_{s,d}$ is double that of case HD5 and the drag-reduction level is larger. An analogous reduction of the number of vortices was observed by \cite{olivucci-ricco-2019} for the flow over actively spinning rings.

\subsection{Flow statistics}
\label{sec:flow-stats}

\begin{figure}
\centering
\hspace{-.0\textwidth}~
\begin{subfigure}[c]{\textwidth}
    \includegraphics[width=1\textwidth]{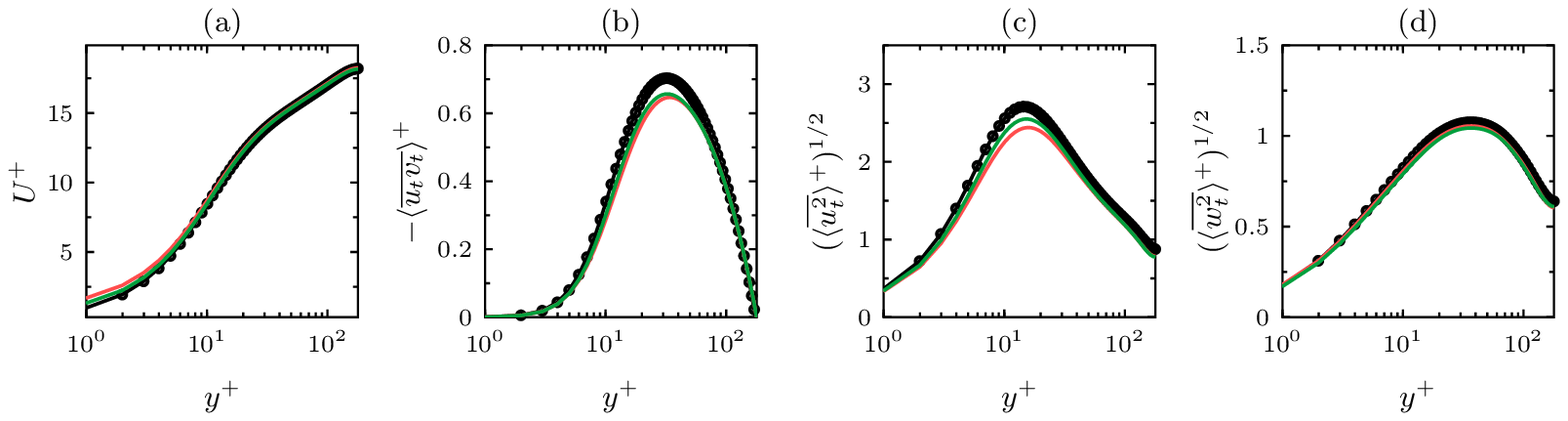}
\end{subfigure}
\caption{
Wall-normal profiles of flow statistics for 
case HD4 (\raisebox{1.5pt}{\includegraphics[height=.1\baselineskip]{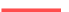}}), 
case HD5 (\raisebox{1.5pt}{\includegraphics[height=.1\baselineskip]{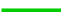}}) 
and the reference channel flow (\includegraphics[height=.5\baselineskip]{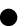}), averaged on the whole wall. The variables are scaled with the wall units of the reference channel flow.
}
\label{fig:stat-whole-half}
\end{figure}

Figure \ref{fig:stat-whole-half} depicts the flow statistics computed by averaging over the entire wall surface. The mean flow is affected in the viscous sublayer, where the gradient is reduced. The peaks of the Reynolds stresses and the root-mean-square of the streamwise velocity are less intense, while the statistics of the wall-normal and spanwise velocities are unvaried.

\begin{figure}
\centering
\begin{subfigure}[l]{\textwidth}
	\includegraphics[width=1\textwidth]{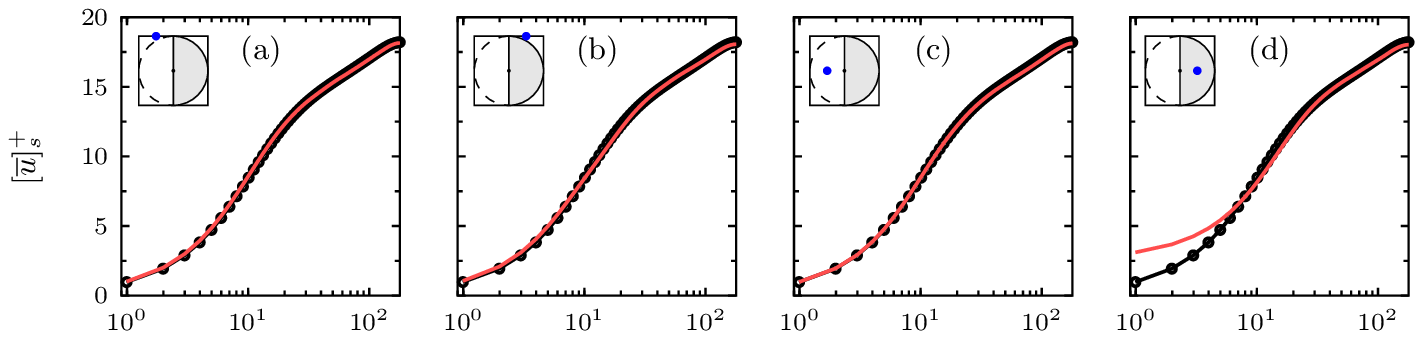}
\end{subfigure}
\begin{subfigure}[l]{\textwidth}
	\includegraphics[width=1\textwidth]{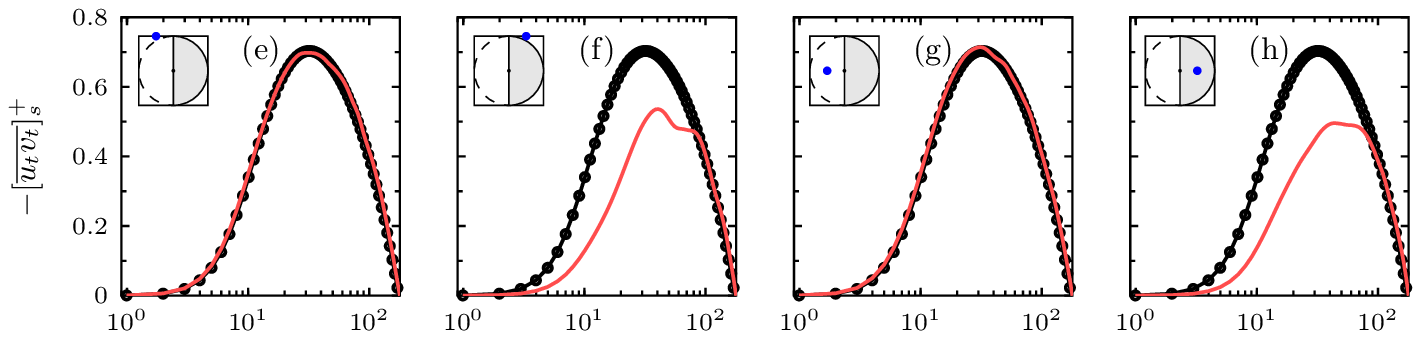}
\end{subfigure}
\begin{subfigure}[l]{\textwidth}
	\includegraphics[width=1\textwidth]{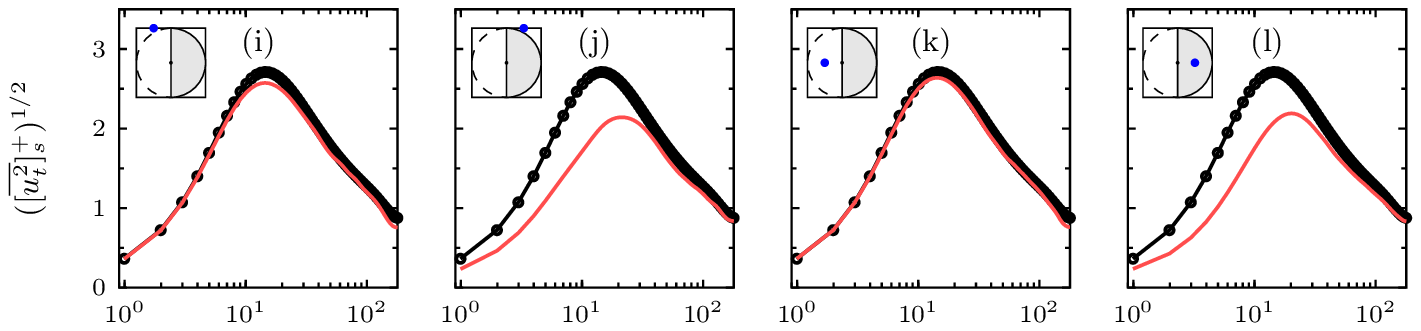}
\end{subfigure}
\begin{subfigure}[l]{\textwidth}
	\includegraphics[width=1\textwidth]{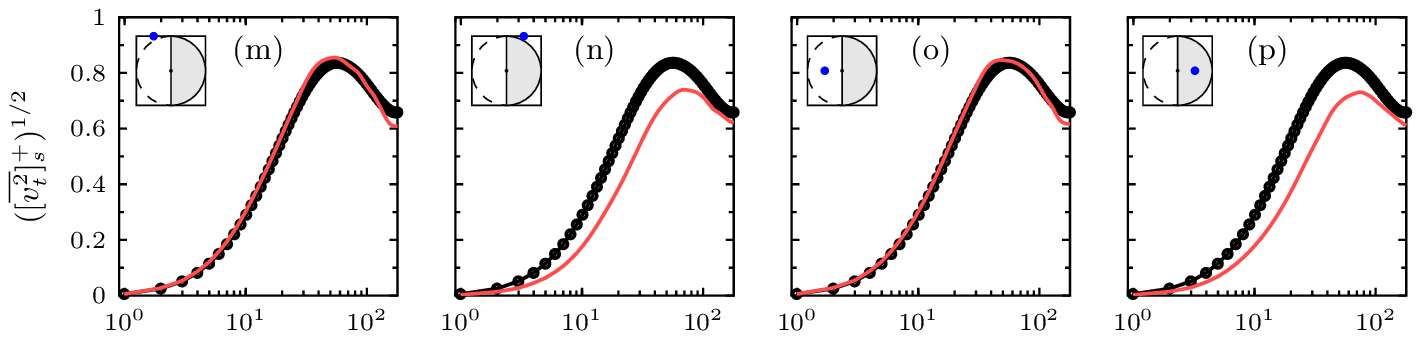}
\end{subfigure}
\begin{subfigure}[l]{\textwidth}
	\includegraphics[width=1\textwidth]{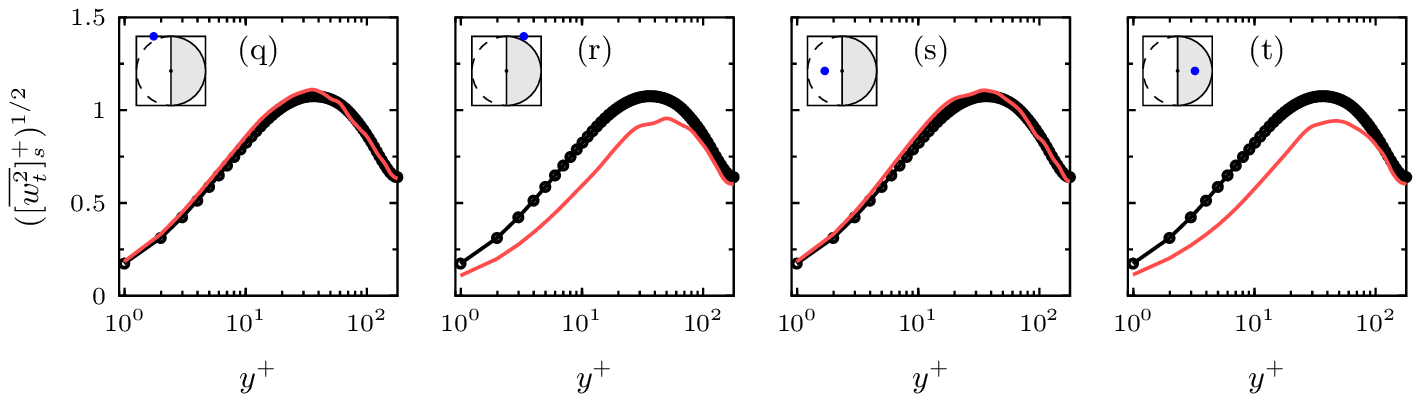}
\end{subfigure}
\caption{
Wall-normal profiles of several flow statistics at four different positions on the wall:
(a)-(d) the period-averaged streamwise velocity $[\overline{u}]_s^+$, 
(e)-(h) the turbulent Reynolds stress $[\overline{u_t v_t}]_s^+$,
(i)-(l) the root-mean-square of the streamwise velocity.
(m)-(p) the root-mean-square of the wall-normal velocity.
(q)-(t) the root-mean-square of the spanwise velocity.
The positions are identified by blue dots in the sketches.
The statistics of case HD4 (\raisebox{1.5pt}{\includegraphics[height=.1\baselineskip]{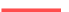}}) are compared to the reference channel-flow statistics (\includegraphics[height=.5\baselineskip]{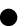}).
The variables are scaled with the wall units of the reference channel flow.
}
\label{fig:stat-pos-half}
\end{figure}

Figure \ref{fig:stat-pos-half} shows the wall-normal profiles of time- and ensemble-averaged statistical quantities at different locations in the proximity of and on the half-disc surface. The streamwise mean velocity $[\overline{u}]_s^+$ is unaffected on the left fixed-wall part and immediately downstream of the disc. It is instead altered significantly up to $y^+=5$ in the middle of the disc surface, where the streamwise slip velocity has the direct effect of reducing the wall-shear stress. In the buffer layer and above, the flow is instead almost unaltered compared with the fixed-wall case. This result gives support to the main assumption used by KK13 and KK14 for estimating the drag reduction, discussed in the Supplementary Material \ref{app:kk}, i.e., the bulk mean flow above the disc coincides with the reference case, so that the boundary-layer thickness remains constant.

On the left part, the statistics are largely unaffected, with the streamwise-velocity root-mean-square decreasing slightly for $y^+>5$ and the Reynolds stresses $[\overline{u_t v_t}]_s^+$ being indistinguishable from the reference case. The wall-normal profiles of all the fluctuating velocities are attenuated in the same way up to $y^+=80$ on and downstream of the disc surface. This behaviour chimes with the flow response to discs and rings spinning at a constant angular velocity, studied by \cite{olivucci-ricco-2019}. They showed that the flow statistics are largely independent of the streamwise location, i.e., the attenuation of the wall turbulence brought about by the disc rotation persists downstream even where the portion of the wall is stationary. Both $[\overline{u_t v_t}]_s^+$ and $[\overline{u_t u_t}]_s^+$ decrease significantly, shifting upward and reducing by a maximum of $30\%$.

\begin{figure}
\centering
\begin{subfigure}[l]{\textwidth}
    \includegraphics[width=1\textwidth]{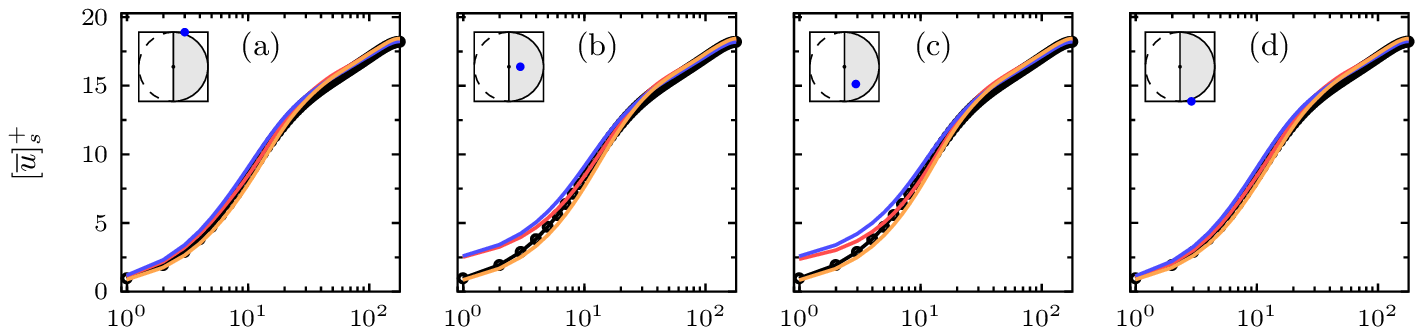}
\end{subfigure}
\begin{subfigure}[l]{\textwidth}
    \includegraphics[width=1\textwidth]{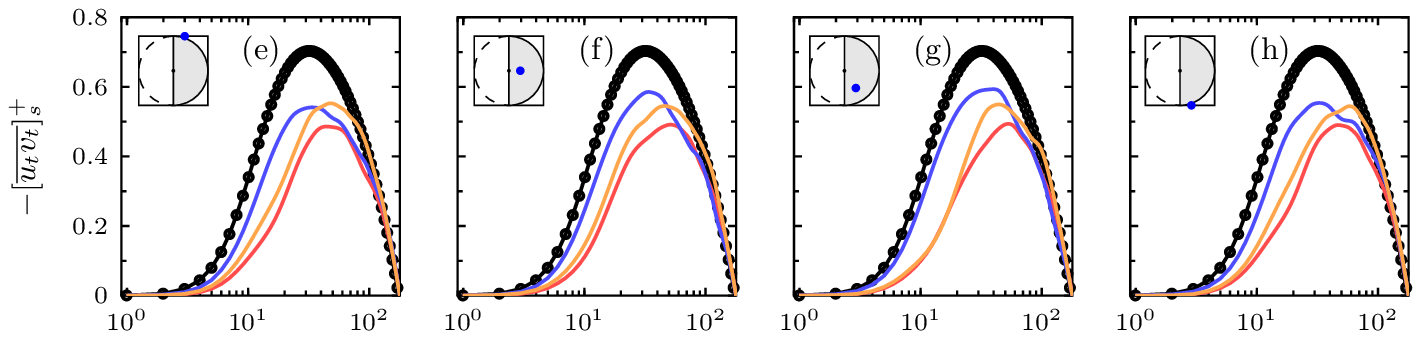}
\end{subfigure}
\begin{subfigure}[l]{\textwidth}
    \includegraphics[width=1\textwidth]{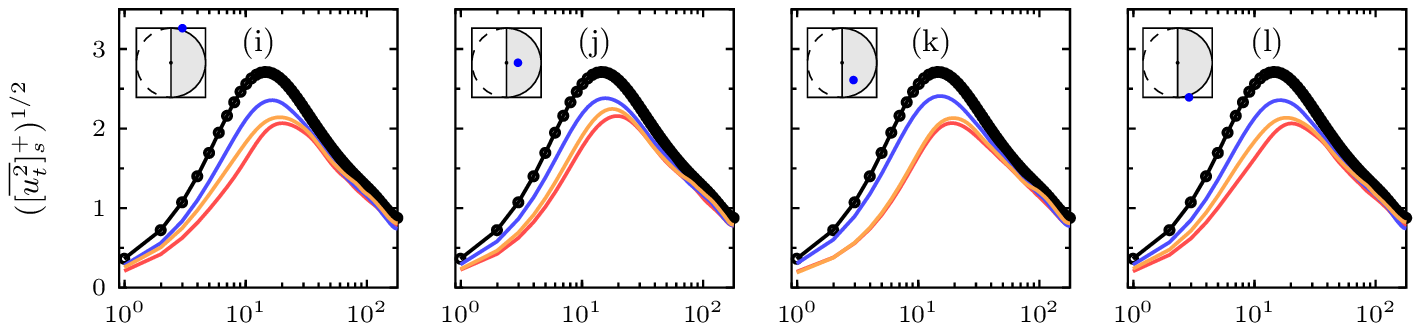}
\end{subfigure}
\begin{subfigure}[l]{\textwidth}
    \includegraphics[width=1\textwidth]{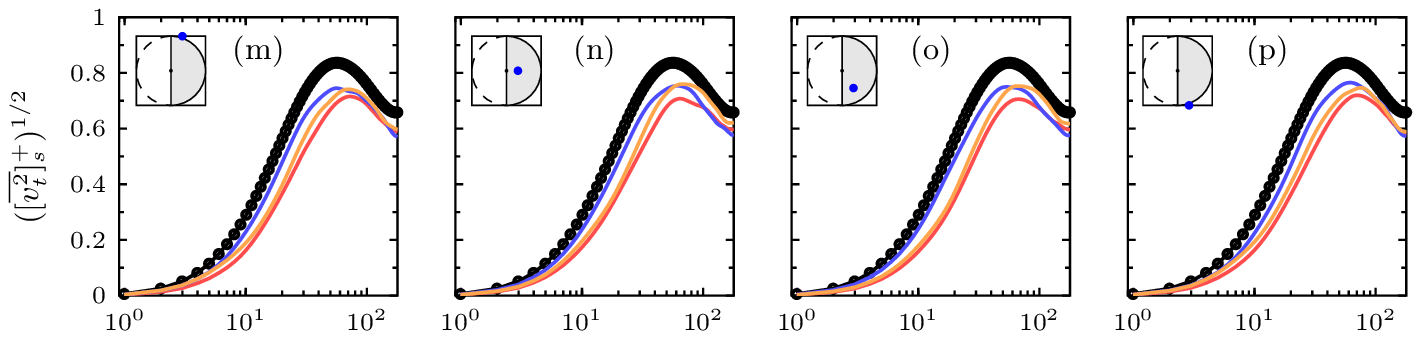}
\end{subfigure}
\begin{subfigure}[l]{\textwidth}
    \includegraphics[width=1\textwidth]{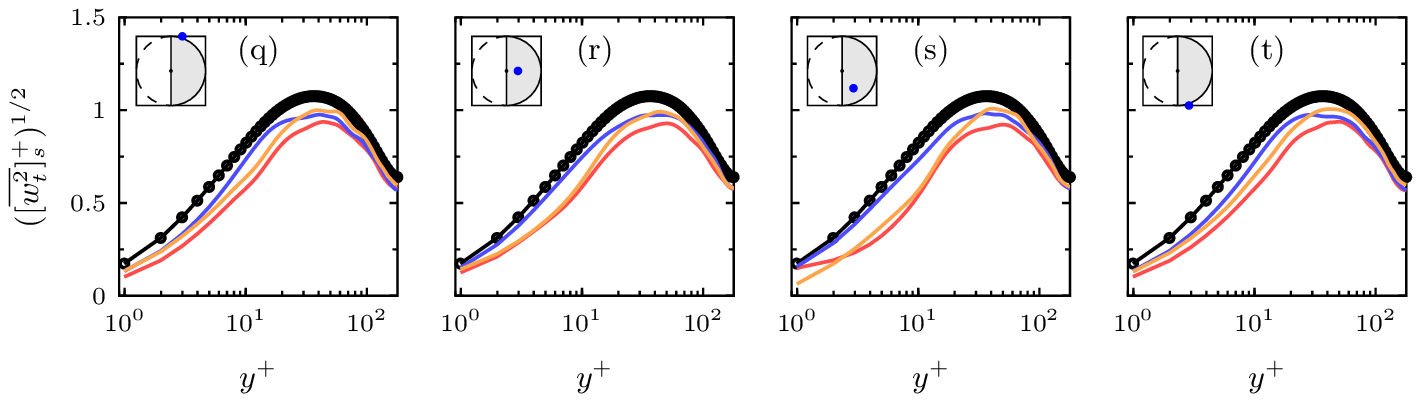}
\end{subfigure}
\caption{
Wall-normal profiles of several period-averaged wall-normal statistics at different positions on the half-disc. The location are shown in the sketches. Case HD4 statistics (\raisebox{1.5pt}{\includegraphics[height=.1\baselineskip]{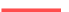}}) are compared to the reference channel-flow (\includegraphics[height=.5\baselineskip]{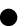}) and to the streamwise-only (\raisebox{1.5pt}{\includegraphics[height=.1\baselineskip]{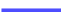}}) and spanwise-only (\raisebox{1.5pt}{\includegraphics[height=.1\baselineskip]{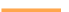}}) test cases. The variables are scaled with the wall units of the reference channel.
}
\label{fig:stat-pos-half-stream-span}
\end{figure}

\begin{figure}
\centering
\begin{subfigure}[l]{.33\textwidth}
\centering
    \includegraphics{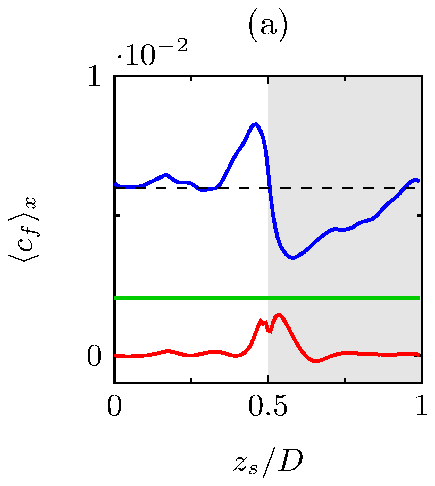}
\end{subfigure}
\hspace{.02\textwidth}
\begin{subfigure}[l]{.3\textwidth}
\centering
    \includegraphics{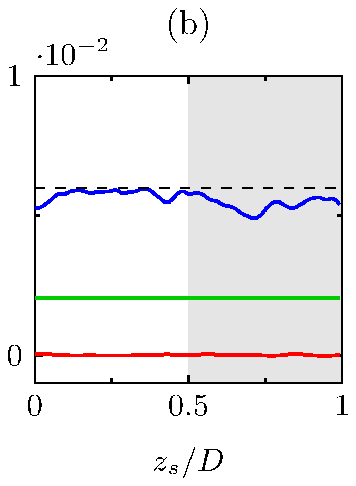}
\end{subfigure}
\begin{subfigure}[l]{.3\textwidth}
\centering
    \includegraphics{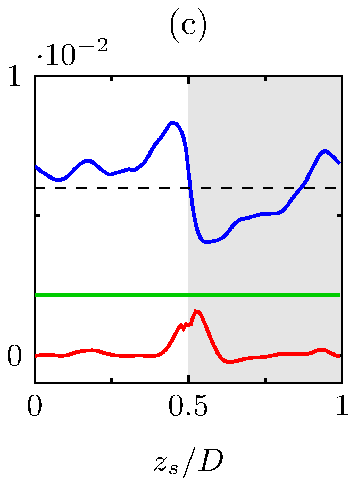}
\end{subfigure}
\caption{
Distribution of the local skin-friction coefficients according to \eqref{eq:fik-passive}.
(a) case HD4, (b) streamwise only, (c) spanwise only.
Budget terms 
$c_f^d$ (\raisebox{1.5pt}{\includegraphics[height=.1\baselineskip]{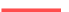}}),
$c_f^t$ (\raisebox{1.5pt}{\includegraphics[height=.1\baselineskip]{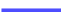}}),
$c_f^{lam}$ (\raisebox{1.5pt}{\includegraphics[height=.1\baselineskip]{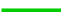}}),
reference channel $c_{f,0}^t$ (\raisebox{1.5pt}{\includegraphics[height=.1\baselineskip]{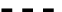}}).
}
\label{fig:fik-curves-half}
\end{figure}

Figure \ref{fig:stat-pos-half-stream-span} reports the flow statistics of the cases with streamwise-only and spanwise-only wall forcing, discussed in \S\ref{sec:analogies}. The statistics are computed on the half disc along a line parallel to the streamwise axis of symmetry of the disc. The line is positioned close to the axis of symmetry because in that region the spanwise motion amplitude is relatively large.  The mean velocity profile of the streamwise-only case and case HD4 are very close, which is expected given that the mean slip velocities are similar.
The local peak intensity of the Reynolds stresses is more weakened by the full case HD4 than either one of the streamwise-only or spanwise-only cases. The latter is however more efficient (at least locally) at pushing the peak of the stresses away from the wall and is also locally more efficient at weakening all the three turbulence intensity components, as shown in figure \ref{fig:stat-pos-half-stream-span}i-t.
The large reductions of turbulence quantities given by the spanwise-only case along this streamwise line do not explain the small drag-reduction margin reported in table \ref{tab:stream-span} when the whole disc surface is considered. We thus resort to the spanwise distributions of the local skin-friction coefficient, based on our version of the Fukagata-Iwamoto-Kasagi identity developed for spinning-disc flows \citep{olivucci-ricco-2019}. The budget equation is
\begin{equation}
\averspace{c_f}_x (z_s) = C^{lam}_f + \averspace{c^d_f}_x (z_s) + \averspace{c^t_f}_x (z_s),
\label{eq:fik-passive}
\end{equation}
where $C^{lam}_f$ is the laminar skin-friction coefficient, $c^d_f$ and $c^d_f$ are the local contributions to $C_f$ from $u_t v_t$ and $u_d v_d$ respectively, and the subscript $x$ denotes averaging along the streamwise direction.
Averaging \eqref{eq:fik-passive} along $z$ gives the budget identity for the global skin-friction coefficient $C_f = C^{lam}_f + C^{d}_f + C^{t}_f$. 
The terms of \eqref{eq:fik-passive} are depicted in figure \ref{fig:fik-curves-half}. In the spanwise-only case, the wall-shear stress is reduced on the disc near the centreline, but it also increases as the right disc edge is approached (blue line in figure~\ref{fig:fik-curves-half}c). These two effects balance each other, causing the low drag reduction. The detrimental contribution of $c_f^d$ is absent when the spanwise component is eliminated, as the flat red curve in figure~\ref{fig:fik-curves-half}b indicates.

\subsection{Power consumed by the moving discs}
\label{sec:pwr}
The passive discs can be classified as a passive-absorbing drag-reduction technique, following the definition first proposed by \cite{aghdam-ricco-2016}, who considered the channel as the control volume for the energy balance. Passive-absorbing methods do not require external power to cause drag reduction, but they consume power as the wall turbulence interacts with the wall, which is fitted by half discs in our case. Another example of passive-absorbing method is compliant surfaces that deform under the action of the wall turbulence. The other group of passive techniques is called passive-neutral, such as riblets \citep{garcia-jimenez-2011} or dimples \citep{vannesselrooij-etal-2016} that instead do not transfer power from the wall turbulence to the exterior of the fluid control volume because they are rigid and stationary. Passive-neutral devices nonetheless involve a power penalty within the fluid control volume with respect to the reference case, caused by a larger wetted area for riblets and dimples, or detrimental pressure drag for dimples, wavy walls \citep{ghebali-etal-2017} and baffles \citep{marensi-etal-2020}.

\begin{figure}
\centering
\tikzset{
  block1/.style = {draw, rectangle,
    minimum height=2.5cm,
    minimum width=4cm},
  block2/.style = {draw, rectangle,
    minimum height=2.5cm,
    minimum width=4cm},
  input_pp/.style = {coordinate},
  input_pm/.style = {coordinate},
  output_eps/.style = {coordinate},
  output_fd/.style = {coordinate},
}
\includegraphics[width=.33\textwidth]{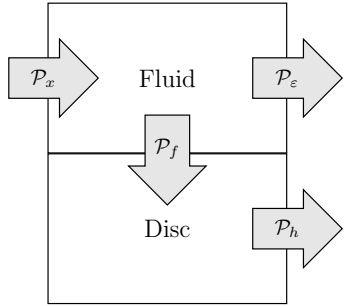}
\vspace{0.5cm}
\caption{
Schematic representation of the mean power budget for the disc-fluid system. The size of the arrows is not proportional to the power amount.
}
\label{fig:pwr_sketch}
\end{figure}

In the half-disc case, the power extracted from the wall turbulence by the discs is consumed through friction effects below the disc surface.
The power $\mathcal{P}_x=C_f U_b^3 D^2$ required to pump the fluid in a single-disc flow unit along the streamwise direction satisfies the balance:
\begin{equation}
    \mathcal{P}_x = \mathcal{P}_\varepsilon + \mathcal{P}_f,
\label{eq:pwr_budget}
\end{equation}
which follows from the integration of the mean kinetic energy equation on the flow-unit volume \citep{hinze-1975}. The power-budget terms are visualized in figure \ref{fig:pwr_sketch} by arrows that represent the power transfer direction.
The power $\mathcal{P}_\varepsilon$ represents the total dissipation in the fluid, i.e., the volume integral of the point-wise viscous dissipation rate. The power $\mathcal{P}_f$ is the power outflow to the moving discs, calculated by integrating the point-wise power on the wall surface, which is equivalent to $\mathcal{P}_f = \overline{T}_f \Omega$ in accordance to \eqref{eq:t_f}, where $\Omega$ is the disc angular velocity.
In the fixed-wall case, the power budget is $\mathcal{P}_{x,un}=\mathcal{P}_{\varepsilon,un}$.
Since the disc rotates steadily, $\mathcal{P}_f = \mathcal{P}_h = 
(\overline{T}_h +T_b) \Omega$, i.e., the power extracted by the disc from the wall turbulence is all dissipated by the frictional losses occurring in the disc housing. In the best drag-reduction case HD4, $\mathcal{P}_\varepsilon$ amounts to $99\%$ of $\mathcal{P}_x$, so $\mathcal{P}_f$ accounts for the remaining $1\%$.

\section{Conclusions}
\label{sec:conclusions}

Turbulent channel flows over discs that rotate passively under the action of the wall-shear stress have been studied via direct numerical simulations. The dynamics of the discs is driven by the instantaneous torque produced by the overriding wall turbulence and the Navier-Stokes equations are solved with modified boundary conditions that synthesize the motion of the discs. They are therefore two-way-coupling simulations of the equations of motion of the discs and the fluid. The disc dynamics is driven actively by the fluid torque and thwarted by two resisting torques, i.e., the cavity torque, that depends on the flow regime occurring in the cavity, and the torque given by the ball bearing that supports a disc. 

We have considered discs of different diameters that are either fully exposed to the wall turbulence or partially covered by the wall. In the full-disc cases, the root-mean-square of the amplitude of the disc velocity is always smaller than the wall-friction velocity. The wall displacement is so small that the minimal forcing conditions needed to alter the wall turbulence are not met and no drag reduction occurs. An optimal diameter of $D^+=100$, matching the characteristic spanwise spacing of the near-wall low-speed streaks, generates the largest disc displacement and velocities. A reduced-order model, discussed in the Supplementary Material \ref{app:uncoupled-dynamics}, has helped clarify the spectral filtering of the wall-turbulence signal to the disc motion.
In the fastest half-disc case, drag-reduction levels up to $20\%$ occur over the disc surface and up to $5.6\%$ when the entire channel walls are considered. We also computed the power extracted by the discs, which is dissipated by the friction effects below the disc surface. The only experimental results on passively-rotating half discs were analyzed \citep{koch-kozulovic-2013,koch-kozulovic-2014}. The drag-reduction margin was not measured in those experiments, but estimated to be $17$\% through a fixed-wall skin-friction correlation. It was therefore not possible to carry out quantitative comparisons. Nevertheless, the experimental trend of drag reduction as a growing function of an equivalent slip velocity matches our numerical data qualitatively and the angular velocities are of comparable amplitude.  Further work is needed to investigate the dependence of the disc-slip velocity and the drag-reduction margin on the Reynolds number. 
\section*{Acknowledgements}

The authors acknowledge the EPSRC Grant No. EP/L000261/1 and the UK Turbulence Consortium for providing computer time. The authors would also like to thank the reviewers for the useful comments, Dr M. R\"{u}tten at DLR in G{\"o}ttingen for providing the technical specifications of his disc-fitted boundary-layer experimental setup, Prof. R. Dwyer-Joyce at the University of Sheffield for the useful discussion on the bearing torque, Prof. N. Qin also at the University of Sheffield for pointing out the experimental study of Koch and Kozulovic to us, and Prof. K. Taira at University of California, Los Angeles for the interesting discussions about the disc response dynamics. PO and DW thank the Department of Mechanical Engineering at the University of Sheffield for providing financial support, and PO thanks the Gordon Franklin Scholarship for sponsoring his Ph.D. studies. PR was partially funded by the H2020 EU ``DRAGY'' project.

\section*{Declaration of Interests}
The authors report no conflict of interest.

\bibliographystyle{jfm}
\bibliography{freely}

\begin{thebibliography}{46}
\expandafter\ifx\csname natexlab\endcsname\relax\def\natexlab#1{#1}\fi
\def\au#1{#1} \def\ed#1{#1} \def\yr#1{#1}\def\at#1{#1}\def\jt#1{\textit{#1}}
  \def\bt#1{#1}\def\bvol#1{\textbf{#1}} \def\vol#1{#1} \def\pg#1{#1}
  \def\publ#1{#1}\def\arxiv#1{#1}\def\org#1{#1}\def\st#1{\textit{#1}}

\bibitem[Aghdam \& Ricco(2016)]{aghdam-ricco-2016}
{\sc \au{Aghdam, S.K.} \& \au{Ricco, P.}} \yr{2016}  \at{Laminar and turbulent
  flows over hydrophobic surfaces with shear-dependent slip length}.  \jt{Phys.
  Fluids}  \bvol{28}~(3),  \pg{035109}.

\bibitem[{\AA}str{\"o}m \& Murray(2008)]{astroem-murray-2008}
{\sc \au{{\AA}str{\"o}m, K.J.} \& \au{Murray, R.M.}} \yr{2008} {\em Feedback
  Systems: An Introduction for Scientists and Engineers\/}.  \publ{Princeton
  University Press}.

\bibitem[Barenghi \& Jones(1989)]{barenghi-jones-1989}
{\sc \au{Barenghi, C.~F.} \& \au{Jones, C.~A.}} \yr{1989}  \at{Modulated
  {{Taylor}}\textendash{{Couette}} flow}.  \jt{J. Fluid Mech.}  \bvol{208},
  \pg{127--160}.

\bibitem[Bechert {\em et~al.\/}(1996)Bechert, Hage \&
  Brusek]{bechert-etal-1996}
{\sc \au{Bechert, D.W.}, \au{Hage, W.} \& \au{Brusek, M.}} \yr{1996}  \at{Drag
  reduction with the slip wall}.  \jt{AIAA J.}  \bvol{34}~(5),
  \pg{1072--1074}.

\bibitem[Blesbois {\em et~al.\/}(2013)Blesbois, Chernyshenko, Touber \&
  Leschziner]{blesbois-etal-2013}
{\sc \au{Blesbois, O.}, \au{Chernyshenko, S.I.}, \au{Touber, E.} \&
  \au{Leschziner, M.A.}} \yr{2013}  \at{Pattern prediction by linear analysis
  of turbulent flow with drag reduction by wall oscillation}.  \jt{J. Fluid
  Mech.}  \bvol{724},  \pg{607--641}.

\bibitem[Busse \& Sandham(2012)]{busse-sandham-2012}
{\sc \au{Busse, A} \& \au{Sandham, ND}} \yr{2012}  \at{Influence of an
  anisotropic slip-length boundary condition on turbulent channel flow}.
  \jt{Phys. Fluids}  \bvol{24}~(5),  \pg{055111}.

\bibitem[Choi(2002)]{choi-2002}
{\sc \au{Choi, K.-S.}} \yr{2002}  \at{Near-wall structure of turbulent boundary
  layer with spanwise-wall oscillation}.  \jt{Phys. Fluids}  \bvol{14}~(7),
  \pg{2530--2542}.

\bibitem[Choi {\em et~al.\/}(1997)Choi, Yang, Clayton, Glover, Atlar, Semenov
  \& Kulik]{choi-etal-1997}
{\sc \au{Choi, K.-S.}, \au{Yang, X.}, \au{Clayton, B.R.}, \au{Glover, E.J.},
  \au{Atlar, M.}, \au{Semenov, B.N.} \& \au{Kulik, V.M.}} \yr{1997}
  \at{Turbulent drag reduction using compliant surfaces}.  \jt{Proc. R. Soc.
  Lond. A.}  \bvol{453}~(1965),  \pg{2229--2240}.

\bibitem[Davidson(2004)]{davidson-2004}
{\sc \au{Davidson, P.A.}} \yr{2004} {\em Turbulence: {A}n {I}ntroduction {F}or
  {S}cientists {A}nd {E}ngineers\/}.  \publ{Oxford University Press}.

\bibitem[{de Giovanetti} {\em et~al.\/}(2016){de Giovanetti}, Hwang \&
  Choi]{degiovanetti-etal-2016}
{\sc \au{{de Giovanetti}, M.}, \au{Hwang, Y.} \& \au{Choi, H.}} \yr{2016}
  \at{Skin-friction generation by attached eddies in turbulent channel flow}.
  \jt{J. Fluid Mech.}  \bvol{808},  \pg{511--538}.

\bibitem[Garc{\'\i}a-Mayoral \& Jim{\'e}nez(2011)]{garcia-jimenez-2011}
{\sc \au{Garc{\'\i}a-Mayoral, R.} \& \au{Jim{\'e}nez, J.}} \yr{2011}  \at{Drag
  reduction by riblets}.  \jt{Phil. Trans. Royal Soc. A}  \bvol{369}~(1940),
  \pg{1412--1427}.

\bibitem[Ghebali {\em et~al.\/}(2017)Ghebali, Chernyshenko \&
  Leschziner]{ghebali-etal-2017}
{\sc \au{Ghebali, S.}, \au{Chernyshenko, S.I.} \& \au{Leschziner, M.A.}}
  \yr{2017}  \at{Can large-scale oblique undulations on a solid wall reduce the
  turbulent drag?}  \jt{Phys. Fluids}  \bvol{29}~(10),  \pg{105102}.

\bibitem[Harris \& Kotzalas(2006)]{harris-kotzalas-2006}
{\sc \au{Harris, T.~A.} \& \au{Kotzalas, M.~N.}} \yr{2006} {\em Essential
  {{Concepts}} of {{Bearing Technology}}\/}.  \publ{{CRC Press}}.

\bibitem[Hinze(1975)]{hinze-1975}
{\sc \au{Hinze, J.O.}} \yr{1975} {\em Turbulence\/}.  \publ{McGraw Hill, Inc.
  -- Second Edition}.

\bibitem[Hu {\em et~al.\/}(2006)Hu, Morfey \& Sandham]{hu-etal-2006}
{\sc \au{Hu, Z.~W.}, \au{Morfey, C.~L.} \& \au{Sandham, N.~D.}} \yr{2006}
  \at{Wall pressure and shear stress spectra from direct simulations of channel
  flow}.  \jt{{AIAA} J.}  \bvol{44}~(7),  \pg{1541--1549}.

\bibitem[Jeong \& Hussain(1995)]{jeong-hussain-1995}
{\sc \au{Jeong, J.} \& \au{Hussain, F.}} \yr{1995}  \at{On the identification
  of a vortex}.  \jt{J. Fluid Mech.}  \bvol{285},  \pg{69--94}.

\bibitem[J{\'o}zsa {\em et~al.\/}(2019)J{\'o}zsa, Balaras, Kashtalyan,
  Borthwick \& Viola]{jozsa-etal-2019}
{\sc \au{J{\'o}zsa, T.I.}, \au{Balaras, E.}, \au{Kashtalyan, M.},
  \au{Borthwick, A.G.L} \& \au{Viola, I.M.}} \yr{2019}  \at{Active and passive
  in-plane wall fluctuations in turbulent channel flows}.  \jt{J. Fluid Mech.}
  \bvol{866},  \pg{689--720}.

\bibitem[Jung {\em et~al.\/}(1992)Jung, Mangiavacchi \& Akhavan]{jung-1992}
{\sc \au{Jung, W.J.}, \au{Mangiavacchi, N.} \& \au{Akhavan, R.}} \yr{1992}
  \at{Suppression of turbulence in wall-bounded flows by high-frequency
  spanwise oscillations}.  \jt{Phys. Fluids A}  \bvol{4}~(8),  \pg{1605--1607}.

\bibitem[von K{\'a}rm{\'a}n(1921)]{vonkarman-1921}
{\sc \au{von K{\'a}rm{\'a}n, T.}} \yr{1921}  \at{{\"U}ber laminare und
  turbulente {{Reibung}}}.  \jt{Z. Angew. Math. Mech.}  \bvol{1}~(4),
  \pg{233--252}.

\bibitem[Kline {\em et~al.\/}(1967)Kline, Reynolds, Schraub \&
  Runstadler]{kline-etal-1967}
{\sc \au{Kline, S.J.}, \au{Reynolds, W.C.}, \au{Schraub, F.A.} \&
  \au{Runstadler, P.W.}} \yr{1967}  \at{The structure of turbulent boundary
  layers}.  \jt{J. Fluid Mech.}  \bvol{30},  \pg{741}.

\bibitem[Koch \& Kozulovic(2013)]{koch-kozulovic-2013}
{\sc \au{Koch, H.} \& \au{Kozulovic, D.}} \yr{2013} Drag reduction by boundary
  layer control with passively moving wall.  \bt{In {\em ASME 2013 Fluids
  Engineering Division Summer Meeting\/}},  \pg{pp. V01BT15A004--V01BT15A004}.
  American Society of Mechanical Engineers.

\bibitem[Koch \& Kozulovic(2014)]{koch-kozulovic-2014}
{\sc \au{Koch, H.} \& \au{Kozulovic, D.}} \yr{2014} Influence of geometry
  variations on the boundary layer control with a passively moving wall.
  \bt{In {\em AIAA SciTech Forum, 52nd Aero. Sc. Meeting\/}},  \pg{p. 0401}.

\bibitem[Laizet \& Lamballais(2009)]{laizet-lamballais-2009}
{\sc \au{Laizet, S.} \& \au{Lamballais, E.}} \yr{2009}  \at{High-order compact
  schemes for incompressible flows: A simple and efficient method with
  quasi-spectral accuracy}.  \jt{J. Comp. Phys.}  \bvol{228},  \pg{5989--6015}.

\bibitem[Laizet \& Li(2011)]{laizet-li-2011}
{\sc \au{Laizet, S.} \& \au{Li, N.}} \yr{2011}  \at{Incompact3d: A powerful
  tool to tackle turbulence problems with up to $\mathcal{O}$($10^5$)
  computational cores}.  \jt{Int. J. Num. Meth. Fluids}  \bvol{67},
  \pg{1735--1757}.

\bibitem[Larsson \& Raven(2010)]{larsson-raven-2010}
{\sc \au{Larsson, L.} \& \au{Raven, H.C.}} \yr{2010} {\em Ship Resistance and
  Flow\/}.  \publ{Society of Naval Architects and Marine Engineers}.

\bibitem[Leschziner {\em et~al.\/}(2011)Leschziner, Choi \&
  Choi]{leschziner-etal-2011}
{\sc \au{Leschziner, M.A.}, \au{Choi, H.} \& \au{Choi, K.-S.}} \yr{2011}
  \at{Flow-control approaches to drag reduction in aerodynamics: progress and
  prospects}.  \jt{Phil. Trans. R. Soc. A}  \bvol{369}~(1940),
  \pg{1349--1351}.

\bibitem[Mahfoze {\em et~al.\/}(2018)Mahfoze, Laizet \&
  Wynn]{mahfoze-etal-2018}
{\sc \au{Mahfoze, O.}, \au{Laizet, S.} \& \au{Wynn, A.}} \yr{2018} Bayesian
  optimisation of intermittent wall blowing for drag reduction of a spatially
  evolving turbulent boundary layer.  \bt{In {\em 10th Int. Conference in
  Computational Fluid Dynamics, Barcelona, Spain\/}}.

\bibitem[Marensi {\em et~al.\/}(2020)Marensi, Ding, Willis \&
  Kerswell]{marensi-etal-2020}
{\sc \au{Marensi, E.}, \au{Ding, Z.}, \au{Willis, A.P.} \& \au{Kerswell, R.R.}}
  \yr{2020}  \at{Designing a minimal baffle to destabilise turbulence in pipe
  flows}.  \jt{J. Fluid Mech.}  \bvol{900}.

\bibitem[Min \& Kim(2004)]{min-kim-2004}
{\sc \au{Min, T.} \& \au{Kim, J.}} \yr{2004}  \at{Effects of hydrophobic
  surface on skin-friction drag}.  \jt{Phys. Fluids}  \bvol{16}~(7),
  \pg{L55--L58}.

\bibitem[van Nesselrooij {\em et~al.\/}(2016)van Nesselrooij, Veldhuis, van
  Oudheusden \& Schrijer]{vannesselrooij-etal-2016}
{\sc \au{van Nesselrooij, M.}, \au{Veldhuis, L.L.M.}, \au{van Oudheusden, B.W.}
  \& \au{Schrijer, F.F.J.}} \yr{2016}  \at{Drag reduction by means of dimpled
  surfaces in turbulent boundary layers}.  \jt{Exp. Fluids}  \bvol{57}~(9),
  \pg{142}.

\bibitem[Olivucci {\em et~al.\/}(2019)Olivucci, Ricco \&
  Aghdam]{olivucci-ricco-2019}
{\sc \au{Olivucci, P.}, \au{Ricco, P.} \& \au{Aghdam, S.K.}} \yr{2019}
  \at{Turbulent drag reduction by rotating rings and wall-distributed
  actuation}.  \jt{Phys. Rev. Fluids}  \bvol{4}~(9),  \pg{093904}.

\bibitem[Orlandi(2012)]{orlandi-2012}
{\sc \au{Orlandi, P.}} \yr{2012} {\em Fluid flow phenomena: a numerical
  toolkit\/}.  \publ{Springer Science \& Business Media}.

\bibitem[Quadrio(2011)]{quadrio-2011}
{\sc \au{Quadrio, M.}} \yr{2011}  \at{Drag reduction in turbulent boundary
  layers by in-plane wall motion}.  \jt{Phil. Trans. Royal Soc. A}
  \bvol{369}~(1940),  \pg{1428--1442}.

\bibitem[Quadrio \& Ricco(2004)]{quadrio-ricco-2004}
{\sc \au{Quadrio, M.} \& \au{Ricco, P.}} \yr{2004}  \at{Critical assessment of
  turbulent drag reduction through spanwise wall oscillations}.  \jt{J. Fluid
  Mech.}  \bvol{521},  \pg{251--271}.

\bibitem[Quadrio \& Ricco(2011)]{quadrio-ricco-2011}
{\sc \au{Quadrio, M.} \& \au{Ricco, P.}} \yr{2011}  \at{The laminar generalized
  {{Stokes}} layer and turbulent drag reduction}.  \jt{J. Fluid Mech.}
  \bvol{667},  \pg{135--157}.

\bibitem[Quadrio {\em et~al.\/}(2009)Quadrio, Ricco \&
  Viotti]{quadrio-ricco-viotti-2009}
{\sc \au{Quadrio, M.}, \au{Ricco, P.} \& \au{Viotti, C.}} \yr{2009}
  \at{Streamwise-travelling waves of spanwise wall velocity for turbulent drag
  reduction}.  \jt{J. Fluid Mech.}  \bvol{627},  \pg{161--178}.

\bibitem[Rastegari \& Akhavan(2018)]{rastegari-akhavan-2018}
{\sc \au{Rastegari, A.} \& \au{Akhavan, R.}} \yr{2018}  \at{The common
  mechanism of turbulent skin-friction drag reduction with superhydrophobic
  longitudinal microgrooves and riblets}.  \jt{J. Fluid Mech.}  \bvol{838},
  \pg{68}.

\bibitem[Reneaux(2004)]{reneaux-2004}
{\sc \au{Reneaux, D}} \yr{2004} Overview on drag reduction technologies for
  civil transport aircraft.  \bt{In {\em European Congress on Computational
  Methods in Applied Sciences and Engineering (ECCOMAS)\/}},  \pg{pp. 24--28}.

\bibitem[Ricco \& Hahn(2013)]{ricco-hahn-2013}
{\sc \au{Ricco, P.} \& \au{Hahn, S.}} \yr{2013}  \at{Turbulent drag reduction
  through rotating discs}.  \jt{J. Fluid Mech.}  \bvol{722},  \pg{267--290}.

\bibitem[Ricco {\em et~al.\/}(2012)Ricco, Ottonelli, Hasegawa \&
  Quadrio]{ricco-etal-2012}
{\sc \au{Ricco, P.}, \au{Ottonelli, C.}, \au{Hasegawa, Y.} \& \au{Quadrio, M.}}
  \yr{2012}  \at{Changes in turbulent dissipation in a channel flow with
  oscillating walls}.  \jt{J. Fluid Mech.}  \bvol{700},  \pg{77--104}.

\bibitem[Ricco \& Quadrio(2008)]{ricco-quadrio-2008}
{\sc \au{Ricco, P.} \& \au{Quadrio, M.}} \yr{2008}  \at{Wall-oscillation
  conditions for drag reduction in turbulent channel flow}.  \jt{Int. J. Heat
  Fluid Flow}  \bvol{29},  \pg{601--612}.

\bibitem[Skote(2011)]{skote-2011}
{\sc \au{Skote, M.}} \yr{2011}  \at{Turbulent boundary layer flow subject to
  streamwise oscillation of spanwise wall-velocity}.  \jt{Phys. Fluids}
  \bvol{23},  \pg{081703}.

\bibitem[Stewartson(1953)]{stewartson-1953}
{\sc \au{Stewartson, K.}} \yr{1953}  \at{On the flow between two rotating
  coaxial disks}.  \jt{Math. Proc. Cambridge Phil. Soci.}  \bvol{49}~(2),
  \pg{333--341}.

\bibitem[Viotti {\em et~al.\/}(2009)Viotti, Quadrio \&
  Luchini]{viotti-quadrio-luchini-2009}
{\sc \au{Viotti, C.}, \au{Quadrio, M.} \& \au{Luchini, P.}} \yr{2009}
  \at{Streamwise oscillation of spanwise velocity at the wall of a channel for
  turbulent drag reduction}.  \jt{Phys. Fluids}  \bvol{21}~(11).

\bibitem[Wise {\em et~al.\/}(2014)Wise, Alvarenga \&
  Ricco]{wise-alvarenga-ricco-2014}
{\sc \au{Wise, D.J.}, \au{Alvarenga, C.} \& \au{Ricco, P.}} \yr{2014}
  \at{Spinning out of control: {W}all turbulence over rotating discs}.
  \jt{Phys. Fluids}  \bvol{26}~(12),  \pg{125107}.

\bibitem[Wise \& Ricco(2014)]{wise-ricco-2014}
{\sc \au{Wise, D.J.} \& \au{Ricco, P.}} \yr{2014}  \at{Turbulent drag reduction
  through oscillating discs}.  \jt{J. Fluid Mech.}  \bvol{746},  \pg{536--564}.

\end{thebibliography}

\title[Reduction of friction drag by passively rotating discs]{Reduction of turbulent skin-friction drag by passively rotating discs - \\ supplementary material}

\author{}
\affiliation{}
\pubyear{}
\volume{}
\pagerange{}

\maketitle
\setcounter{section}{0}
\setcounter{table}{0}
\setcounter{figure}{0}
\section{Estimation of the skin-friction drag reduction by \cite{koch-kozulovic-2013,koch-kozulovic-2014}}
\label{app:kk} 

We discuss the procedure adopted by KK13 and KK14 to estimate the turbulent drag reduction on the surface of their passively spinning discs, shown in figure \ref{fig:setups}. 
KK13 and KK14 carried out seven experiments in a closed-circuit wind tunnel. The main flow parameters and results are summarized in table \ref{tab:koch1}.

\begin{table}
\centering
\caption{Flow parameters and results of the boundary-layer experiments from KK13 and KK14.}
\label{tab:koch1}
\begin{tabular}{@{}llllllll@{}}
\hline
$U_\infty^*$ [m/s] & $\delta^*_{99}$ [mm] 	& $f^*$ [1/s] & $Re_\tau$ & $D^+$ & $\overline{W}^+$   & $U_{s,D}^+$ & $\mathcal{R}_d(\%)$   \\ 
\hline
20.32	& 3.98 & 9.94	& 276 &	6943	& 3.00	& 1.27  &	11.2  \\ 
25.38	& 3.80 & 14.33	& 322 &	8481	& 3.54	& 1.50  &	12.6  \\ 
30.41	& 3.67 & 19.01	& 366 &	9980	& 3.99	& 1.69  &	13.9  \\ 
35.6	& 3.55 & 23.78	& 409 &	11500	& 4.33	& 1.84  &	14.7  \\ 
40.7	& 3.46 & 28.78	& 449 &	12973	& 4.65	& 1.97  &	15.6  \\ 
45.86	& 3.38 & 34.04	& 488 &	14444	& 4.94	& 2.09  &	16.3  \\ 
50.96	& 3.31 & 39.04	& 525 &	15882	& 5.15	& 2.19  &	17.1  \\ 
\hline
\end{tabular}
\end{table}

Although KK13 and KK14 measured the streamwise velocity profiles on and downstream of the disc, the wall-shear stress on the disc surface was not measured directly. 
They measured $f^*$, the average number of revolutions per second of the disc, and derived the spatial distribution of the streamwise slip length by assuming that the disc angular was constant, i.e., $u_s^*(z^*) = 2 \pi f^* z^*$, where $z^*$ is the spanwise distance from the disc center.
It was further assumed that the $99\%$ boundary-layer thickness $\delta^*_{99}$ over the disc was identical to the fixed-wall case. The boundary-layer thickness was measured experimentally at three spanwise distances from the disc centre and was in good agreement with the fixed-wall value, supporting the assumption. The wall-shear stress $\tau^*_w$ on the disc surface was estimated by modifying correlations for the skin-friction coefficient and the boundary-layer thickness, valid in the fixed-wall case:

\begin{align}
	\tau^*_w(x^*,z^*) &= 0.0225 \rho^* \nu^{*1/4} \frac{[ U_\infty^*-u_s^*(z^*) ]^{7/4}}{\delta_{99}(x^*)^{1/4}}, 
	\label{eq:kk_shearstress} 
	\\
	\delta^*_{99}(x^*) &= 0.37 (x^* - x^*_v) \left[ \frac{\nu^*}{U^*_\infty (x^* - x^*_v) } \right]^{1/5},
	\label{eq:kk_delta99}
\end{align}
where $U_\infty^*$ is the free-stream velocity, $\nu^*$ is the kinematic viscosity of air, and $x^*_v$ is the virtual origin of the turbulent boundary layer, calculated by upstream extrapolation of the downstream measurements of $\delta^*_{99}$. The spatial distribution of the reduction of the wall-shear stress, $\mathcal{R}_{xz}$, is obtained by use of \eqref{eq:kk_shearstress} with and without the local slip velocity $u_s^*$:
\begin{equation}
    \mathcal{R}_{xz}(\%)=100\left[ 1 - \left( 1- \frac{u_s^*}{U_\infty^*}\right)^{7/4}\right].
    \label{eq:kkrxz}
\end{equation}
The spatially-averaged drag reduction over the half disc, $\mathcal{R}_d$, is computed by numerically integrating \eqref{eq:kkrxz} for our friction Reynolds number $Re_\tau$=180. The result is shown in figure~\ref{fig:lplus_vs_dr}.
\section{Modelling of torques below the disc surface}
\label{app:torques}

The modelling of the torques given by the ball bearing and by the fluid friction in the housing cavity is presented.

\subsection{Torque produced by the ball bearing}

Each disc is supported by a ball bearing fixed onto its shaft, as shown in dark grey in figure \ref{fig:housing}. The resisting torque $T_b$ of a rolling-element bearing is caused by a combination of phenomena, such as lubrication, material deformation, thermal losses, and others. Its complete modelling is complex and the industrial estimates are usually performed through the use of empirical formulas \citep{harris-kotzalas-2006}. The main contribution to $T_b$ is the load-induced rolling friction arising from the pressure contact between the rotary elements and the metal grooves, often named races. The interaction is similar to that experienced by train wheels on rail tracks. 

Our estimate of $T_b$ is based on the empirical formulas provided by a manufacturer of rolling-element bearings, the Swedish company SKF (\url{www.skf.com}). A realistic bearing model must be selected by considering the water-channel flow presented in table \ref{tab:waterchan}. The critical aspect is the capability of the bearing to support the axial load of the weight of the disc without generating a high-friction torque. Watertight seals are often used to protect the lubricant, but they are not modelled in our case because lubrication is not required for our slowly rotating and lightly loaded bearings.

A sound choice is a SKF angular-contact thrust bearing belonging to the series 7009. These bearings produce small friction under high axial loads and are available in sizes that are compatible with the required design. Using the bearing data-sheet parameters and an axial load corresponding to the weight of a disc of diameter $D=5$, the SKF formulas return a friction torque that, translated into our outer units, is $\vert T_b \vert = 0.0001$. The effect of the axial load can be neglected because the weight of the disc is negligible. The use of a bearing torque that does not depend on $W$ is justified because, at our slow rotation rates, only rolling friction occurs, which is nearly independent of the angular velocity. The bearing torque is therefore modelled as $T_b = \vert T_b \vert {\rm sgn}(W)$. The sign of the disc-tip velocity is accounted for because the friction of the bearing opposes the rotation.

\subsection{Torque produced by the fluid friction in the housing cavity}
\label{sec:housing_torque}

The motion of the fluid within the cavity under a disc results in a resisting torque $T_h$ acting on the bottom surface of the disc. Two predictive models for $T_h$ are used, thus avoiding the full simulation of the cavity flow.

We consider an idealized case where the disc angular velocity fluctuates following a monochromatic sinusoidal wave of zero mean and frequency $f^*$. 
We assume that the scaled frequency is small, i.e., $d^*_h \sqrt{f^*/\nu^*} \ll 1$, in order to model the cavity flow in a quasi-steady regime (a similar definition for a different oscillating-flow configuration is found in \citealp{barenghi-jones-1989}). In the case of full discs, shown in figure \ref{fig:discs}a, the discs fluctuate randomly, i.e., the time history of the disc motion shows a continuous frequency spectrum. The quasi-steadiness assumption is valid in all the cases because a large fraction of the disc kinetic energy corresponds to a range of frequencies $f^*$ of the spectrum for which  $d^*_h \sqrt{f^* / \nu^*} \ll 1$. In the half-disc cases, shown in figure \ref{fig:discs}b, the discs rotate with a finite mean angular velocity and thus the quasi-steadiness assumption is also always verified due to the standard deviation of the velocity fluctuations being much smaller than their temporal average. A further assumption is that the flow in the cavity is laminar. The steady flow between two coaxial infinite discs, one at rest and the other one moving with constant angular velocity $\Omega$, was studied by \cite{stewartson-1953} in the laminar regime at small and moderate Reynolds numbers. The azimuthal velocity $u_\theta$ is independent of the other two components, $u_\theta =  r \Omega y_h/d_h$, where $y_h$ is the wall-normal coordinate within the cavity and $r$ is the radial coordinate with origin at the centre of the disc. The resisting torque is:
\begin{equation}
	T_h = \frac{\pi D^4 }{32 d_h Re_p}\Omega.
\label{eq:th_stewartson}
\end{equation}
We also model the flow in the cavity through the swirling boundary-layer solution of \cite{vonkarman-1921} for the flow induced by a disc beneath a semi-infinite fluid. The torque of the von-K{\'a}rm{\'a}n solution reads:
\begin{equation}
	T_h = \mbox{sgn}(\Omega) \frac{\pi G D^4 }{32 Re_p^{1/2} } \vert \Omega \vert^{3/2},
\label{eq:th_vk}
\end{equation}
where $G=0.6159$ is a numerically-determined constant.  

The most rigorous method to check the validity of the assumptions would be to carry out a DNS simulation of the cavity flow for each case, from which the scaled frequency parameter can be computed and the mean profiles and the resisting torque can be checked against the predictive model. This approach was not pursued because it is too computationally demanding. Future work should certainly be directed at improving the predictive model of the resisting cavity torque, mainly in view of quantitative comparisons with experimental data. Additional factors that will have to be modelled include the viscous shear stresses on the shaft supporting the disc, the secondary flows inside the cavity due to the ball bearing and the uneven steps or gaps, such as those of the experimental set-up shown in figure \ref{fig:setups}, and the effects of the transitional and turbulent flows inside the cavity at large angular velocities. The fluid friction below the disc could be minimized by designing the disc housing to be air filled, thus sealed from the turbulent flow of water on top of the disc.
\section{Uncoupled dynamics of the full discs}
\label{app:uncoupled-dynamics} 

\subsection{Modelling of the uncoupled dynamics of the full discs}
\label{app:model-uncoupled-dynamics} 

\begin{figure}
\centering
\includegraphics[width=1\textwidth]{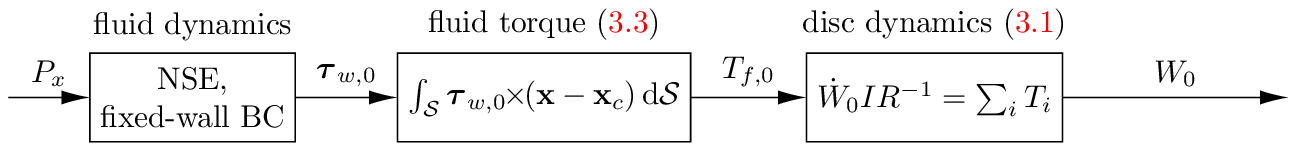}
\caption{Block-diagram of the uncoupled disc-fluid system.}
\label{fig:uncoupled}
\end{figure}

The full-disc configuration is also studied under the simplifying assumption that the disc-tip velocity $W$ is small. This assumption is reasonable as verified by the two-way-coupling simulations. In this uncoupled system, the torque engendered by the wall turbulence causes the motion of the discs, but the disc-tip velocity is so small that the wall turbulence is unaffected by the disc rotation. It is therefore a one-way coupling system, where the NSE are subject to the no-slip stationary-wall boundary conditions. Figure \ref{fig:uncoupled} depicts the uncoupled system, where there is no feedback from the disc to the fluid boundary conditions. The wall-shear-stress torque on the disc is first computed using the reference fixed-wall turbulent channel flow, and then the torque determines the disc dynamics.

The uncoupled system is obtained by a regular perturbation expansion for small $W$. We define the small parameter $\epsilon=\max_t\left[W^*(t^*)\right]/U_p^*\ll1$ and we further assume that $W/T_f=\mathcal{O}(1)$, which is also reasonable because the instantaneous torque, $T_f\ll1$, is given by the statistically homogeneous wall-shear stress acting over a disc without a preferential angular direction. We first expand the disc velocity as $W(t) = \epsilon W_0^m(t) + \mathcal{O}(\epsilon^2)$ (where $W_0^m=W^*/\max_t\left[W^*\right]=\mathcal{O}(1)$), the fluid velocity and pressure as $(\boldsymbol{u},p)(\mathbf{x},t;W) = (\boldsymbol{u_0},p_0)(\mathbf{x},t) + \epsilon (\boldsymbol{u_1},p_1)(\mathbf{x},t) + \mathcal{O}(\epsilon^2)$, the wall-shear stress as $\boldsymbol{\tau_w}(\mathbf{x},t;W) = \boldsymbol{\tau_{w,0}}(\mathbf{x},t) + \epsilon \boldsymbol{\tau_{w,1}}(\mathbf{x},t) + \mathcal{O}(\epsilon^2)$, and the torque as $T_f(t;W) = \epsilon T^m_{f,0}(t) + \epsilon^2 T^m_{f,1}(t) + \mathcal{O}(\epsilon^2)$. (The subscript ``0'' henceforth indicates quantities that refer to the uncoupled case.)
By substituting these expansions into \eqref{eq:continuity-cartesian-vector}, \eqref{eq:navier-stokes-cartesian-vector}, \eqref{eq:sys1}, and \eqref{eq:disc_bc}, one finds that $\boldsymbol{u_0}$ satisfies the reference fixed-wall NSE equations at leading order and that, at the next order $\mathcal{O}(\epsilon)$, the disc-tip velocity $W_0^m$  satisfies the dynamical equation driven by the fixed-wall torque $T^m_{f,0}$. At this next order, $\boldsymbol{u_1}$ satisfies linearized NSE equations, where the convection is driven by $\boldsymbol{u_0}$ and the wall boundary conditions synthesize the effect of the disc rotation on the fluid motion. We do not solve for $\boldsymbol{u_1}$ as we are interested in the leading-order behaviour. The feedback loop thus vanishes as the leading-order expansion leads to the no-slip stationary-wall boundary conditions. 

The uncoupled model has a number of advantages over the coupled model. First, the uncoupled results can be compared to those from the coupled simulations, assessing the impact of the two-way coupling on the disc dynamics and the fluid flow. Second, in the uncoupled case, the equation of motion of the disc is linear because the torque is a known function of the turbulent flow. Therefore, the system is more easily studied in the frequency domain and the linearity of the disc equation of motion allows the explicit calculation of the frequency response of the disc dynamics to the fluid torque. Third, contrary to the coupled case, it is not required to perform new simulations for cases with discs of different diameter, but it is sufficient to simulate the fixed-wall channel flow once and then, from this flow, extract the fluid torques exerted on the disc-shaped patches of choice. This approach allows the study of a large number of cases: we have considered seventy diameters in the range $D^+=20-1200$.

\subsection{Uncoupled dynamics in the physical domain}
\label{app:uncoupled-physical}

This section discusses the uncoupled disc-fluid dynamics, represented schematically in the block-diagram of figure \ref{fig:uncoupled}. The dynamics is still described by \eqref{eq:dyn_lin}, but it becomes linear in the limit of small $W$ because the torque $T_{f,0}$ is given by the fixed-wall reference flow. The solution to the uncoupled \eqref{eq:dyn_lin} is
\begin{equation}
W_0(t) = 
\frac{16 \text{e}^{-C_h t}}{\pi b \rho_d D^3} \int_0^t \left[T_{f,0}\left(\hat{t}\right) - T_b\right] \text{e}^{C_h \hat{t}}  \, \text{d}\hat{t} ,
\label{eq:dyn_lin_uncpld_exact}
\end{equation}
where $C_h=\left(Re_p d_h b \rho_d\right)^{-1}$ and $W_0(t=0)=0$. 

The red symbols in figure \ref{fig:rms_Ds} show that $W_{0,\mbox{rms}}$ behaves qualitatively as $W_{\mbox{rms}}$ in the coupled-dynamics case, i.e., it decays as $D^{-0.7}$ after the maximum response for $D^+=100$. The fixed-wall fluid torque $T_{f,0}$ also follows a similar behaviour to the coupled case. The dependence of $T_{f,0}$ on $D$ in the large-$D$ range matches that of the coupled case, i.e., it also grows as $D^{2.2}$. The values of $W_{0,\mbox{rms}}$ and $T_{f,0,\mbox{rms}}$ are up to three times larger than in the coupled case, regardless of $D$. This result proves that the two-way interaction between the disc and the fluid produces an attenuating effect on the disc velocity and the torque. 

The time evolutions of the fluid torque components $T^x_{f,0}$ and $T^z_{f,0}$ for the uncoupled case are shown in figure \ref{fig:tf_components}a (bottom). Differently from the coupled case, where the torque components are strongly anti-correlated, there is no noticeable correlation between $T^x_{f,0}$ and $T^z_{f,0}$. The contributions of the components of equation \eqref{eq:variance_decomp} to Var$(T_{f,0})$ for the uncoupled case are presented in figure~\ref{fig:tf_components}b by the blue symbols. Similar to the coupled simulations, the streamwise component of the torque dominates over the other two components. The magnitudes of the variance $\avertime{T_{f,0}^{z 2}}$ and covariance $\avertime{T_{f,0}^x T_{f,0}^z}$ are much smaller than in the coupled simulations. 

\begin{figure}
    \centering
	\begin{subfigure}[h]{\textwidth}
	\centering
	\includegraphics[width=1\textwidth]{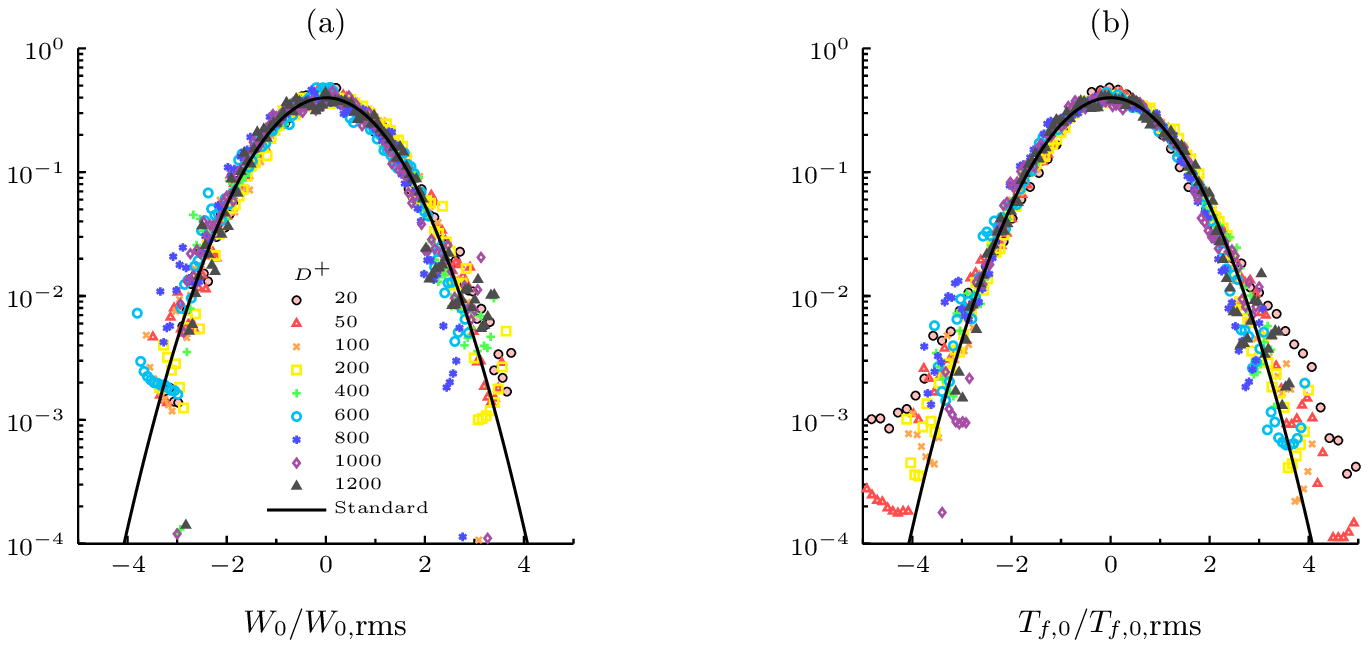}
	\end{subfigure}
    \caption{Standardized histogram plots of (a) $W_0$ and (b) $T_{f,0}$ for uncoupled-dynamics simulations. The solid line denotes the standard normal distribution.}
    \label{fig:pdf_Ds_uncpld}
\end{figure}

Figure \ref{fig:pdf_Ds_uncpld} shows the standardized histograms of $W_0$ and $T_{f,0}$. The coupling has a negligible effect on the trends as the $W_0$ values follow the normal distribution for all the diameters and the heavy-tail behaviour of $T_{f,0}$ deviates from the normal curve at small diameters by a similar amount as in the coupled case. 

\subsection{Uncoupled dynamics in the frequency domain}
\label{app:uncoupled-frequency}

\begin{figure}
\includegraphics[width=1\textwidth]{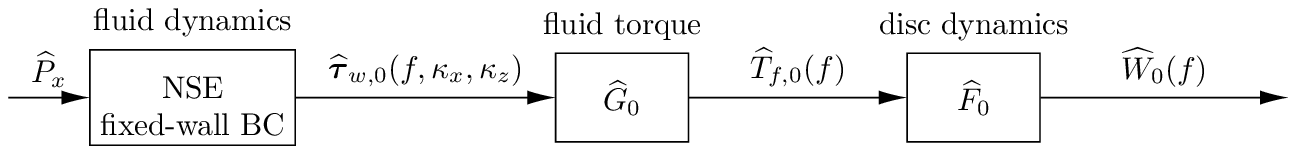}
\vspace{0.1cm}
\caption{Block-diagram of the uncoupled disc-fluid system in the wavenumber-frequency domain, where $\widehat{\boldsymbol{\tau}}_{w,0}(\kappa_x,\kappa_z,f)$ is the wavenumber-frequency spectrum of the wall-shear stress, $\kappa_x$ and $\kappa_z$ are the streamwise and spanwise wavenumbers, $f$ is the frequency, and the hat indicates the Fourier transform. $\widehat{G}_0$ and $\widehat{F}_0$ are the Fourier transforms of the fluid-torque integral \eqref{eq:t_f} and the disc dynamical equation \eqref{eq:dyn_lin}, respectively. 
}
\label{fig:loop_spec}
\end{figure}

The dynamics of the full discs for small $W$ is also investigated in the frequency domain, as shown in figure \ref{fig:loop_spec}. As the response functions are difficult to study in the nonlinear coupled case, we focus on the linear uncoupled system. 

\begin{figure}
   \centering
   \begin{subfigure}[h]{\textwidth}
   \centering
   \includegraphics[width=1\textwidth]{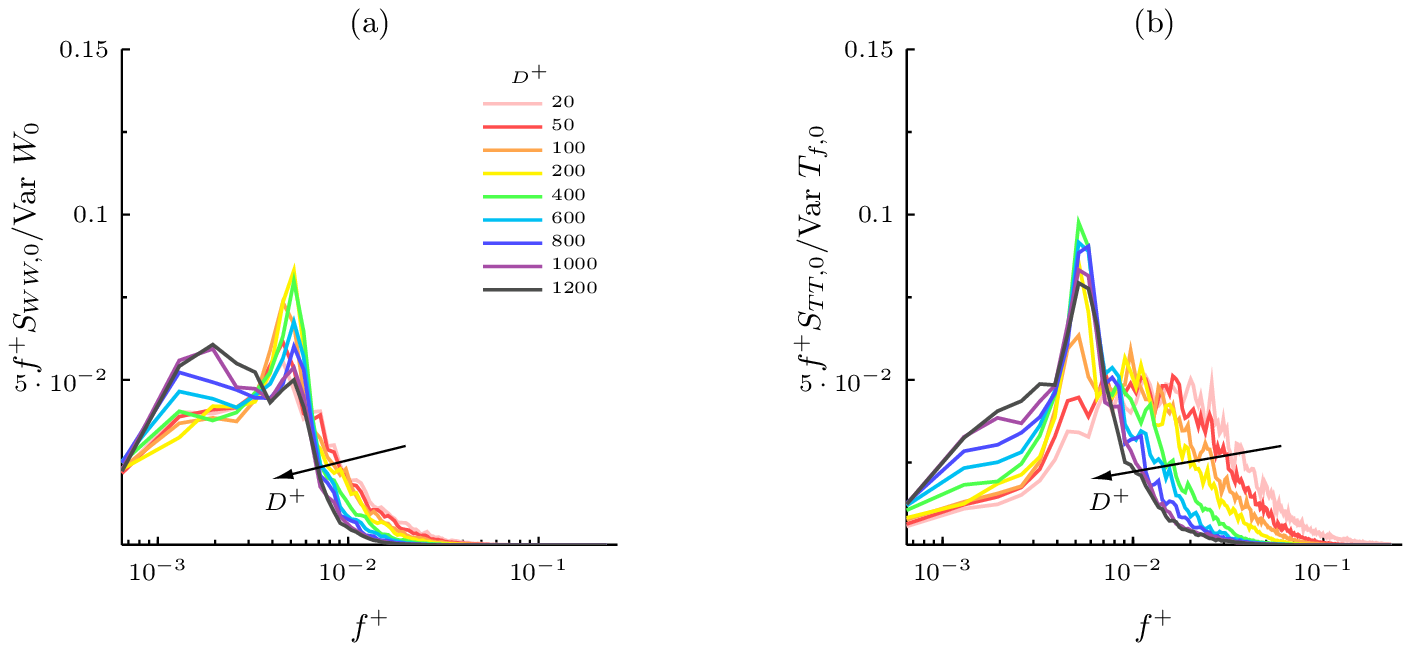}
    \end{subfigure}
    \caption{
    Power spectral densities, denoted by $S$ with subscripts indicating the quantity and shown in pre-multiplied form. (a) Disc-tip velocity $W_0$ and (b) fluid torque $T_{f,0}$, computed from uncoupled-dynamics simulations. The plots are normalized with the total power.}
    \label{fig:psd_Ds_scaled_uncpld}
\end{figure}

The normalized PSDs of the disc-tip velocity and the fluid torque are displayed in figure \ref{fig:psd_Ds_scaled_uncpld}. The PSDs of $W_0$ also follow an analogous behaviour to the coupled case, with the lower frequencies contributing more to the total power as $D$ increases. Similar to the coupled case, the trends shift to low frequencies as the diameter increases and both quantities show the peak at about $f^+$=0.005. These maxima are now much more distinct than in the coupled case and the energy distribution of $T_{f,0}$ at higher frequencies is flatter.  

\begin{figure}
    \centering
    \begin{subfigure}[h]{.47\textwidth}
	\centering
 	  \includegraphics[width=1\textwidth]{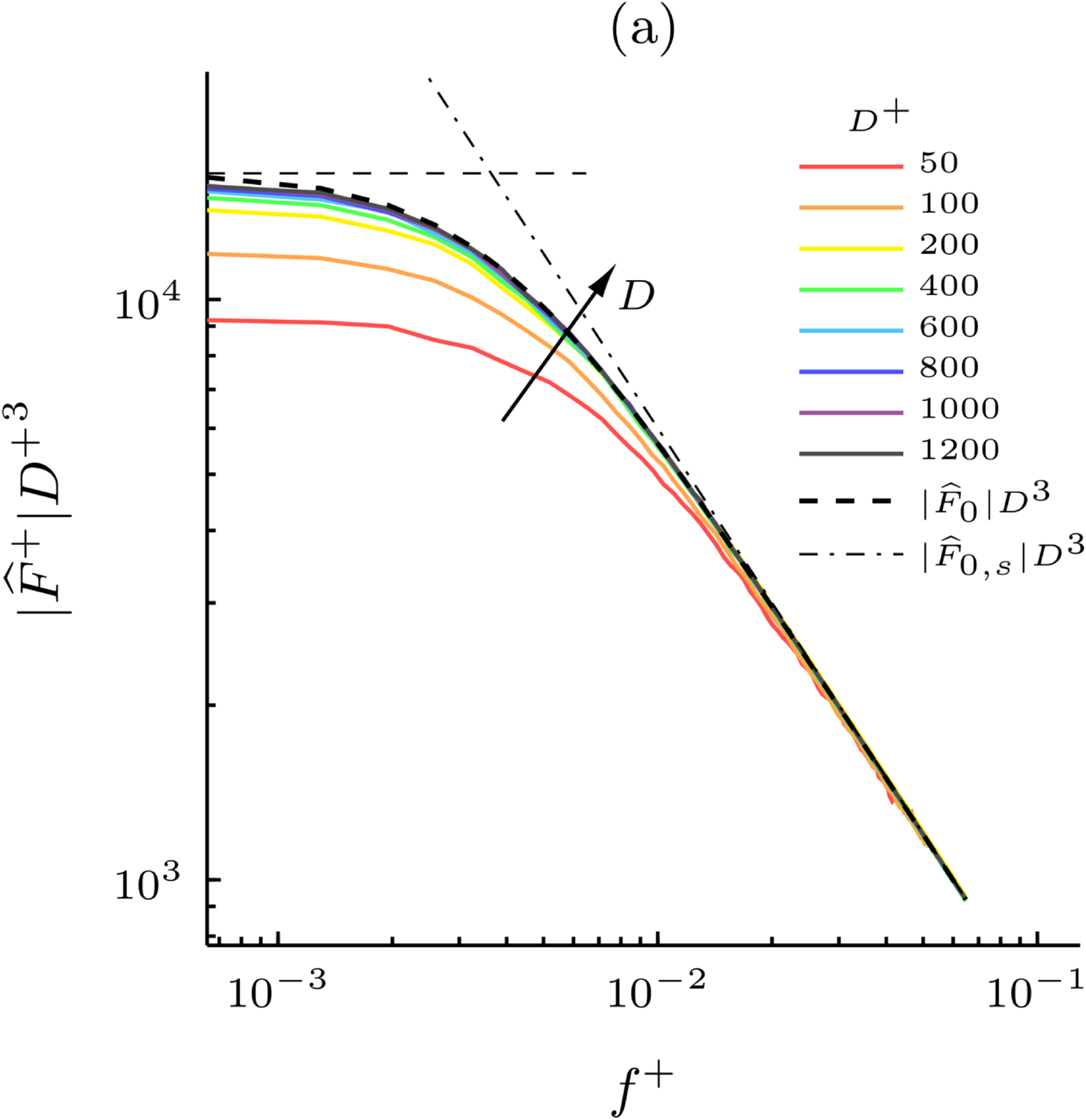}
	\end{subfigure}
	\hspace{.05\textwidth}
	\begin{subfigure}[h]{.45\textwidth}
	\centering
	   \includegraphics[width=1\textwidth]{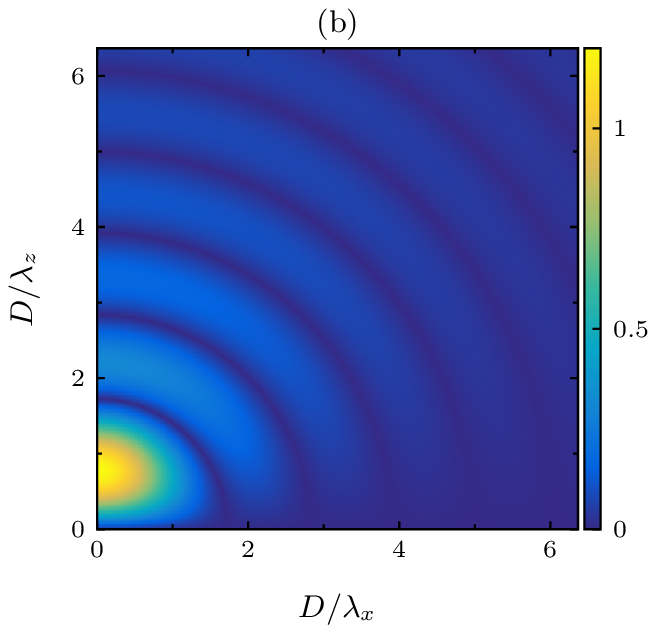}
	\end{subfigure}
    \caption{
        (a) Transfer functions from the uncoupled exact theory (dashed lines) and calculated from the coupled DNS simulations (solid lines). (b) Colour map plot of $\vert \Psi^x_R \vert$ as a function of the inverse of the streamwise and spanwise wavelengths, normalized with the disc diameter. This choice of coordinates emphasizes how the relative scale of the shear-stress wavelengths to the disc size affects the filtering properties of $\Psi^x_R$. 
    }
    \label{fig:reduced_models}
\end{figure}

The frequency-domain solution of \eqref{eq:sys1} is:
\begin{equation}
\widehat{W}_0 = \frac{16}{\pi b \rho_s D^3 (2 \pi i f  + C_h) } \widehat{T_{f,0}}, 
\label{eq:dyn_transf}
\end{equation}
where we have excluded $T_b$ because $T_b \ll T_{h,0}$, revealing the scaling with $D^{-3}$ and the inverse dependence on $b$ and $\rho_s$, i.e., a thinner disc or a lighter material result in larger $W$.
Equation \eqref{eq:dyn_transf} can be written as $\widehat{F}_0= \widehat{W}_0 / \widehat{T_{f,0}}$, so that the gain is $\vert \widehat{F}_0 \vert = 16/\pi b \rho_s D^3 \sqrt{ 4 \pi^2 f^2  + C_h^2 }.$
By neglecting friction to extract information about inertia, $\vert \widehat{F}_{0,s} \vert = 16/\pi b \rho_s D^3 2 \pi f$. Figure \ref{fig:reduced_models} depicts the scaled $\vert \widehat{F} \vert$. When computed from the coupled nonlinear simulations, the procedure is not rigorous, but it offers a tool to assess whether the model is valid. The dashed lines show the matching with the inertia-driven behaviour for large frequencies $\sim f^{-1}$ and the constant behaviour at low frequencies, dictated by the housing-torque viscous effects. We find $\widehat{F}_0 \geq \widehat{F}$ at any $f$, with the difference reducing as $D^{-0.8}$ as $D$ increases.

To study $\widehat{G}_0$, shown in figure \ref{fig:loop_spec}, we neglect the dependence of $T_{f,0}$ on time and study its variance using temporally uncorrelated realizations $T_{f,0}(t_i)$, where the $t_i$ are sufficiently delayed. We study the $x$-component $T_{f,0}^x$, given in \eqref{eq:t_f} (the analysis is analogous for $T_{f,0}^z$), i.e.,
\begin{align}
    T_{f,0}^x(t_i; R)  
    =&
    \int_{\mathcal{S}} z \mathcal{F}^{-1}_{xz} \big\{\widehat{\tau}_{w,x,0}\big\} (\kappa_x, \kappa_z) \} \mbox{d}x \mbox{d}z= 
    \\
    =&
    \iint_{\mathbb{R}^2} \widehat{\tau}_{w,x,0}  (\kappa_x, \kappa_z) 
    \left( \frac{1}{4\pi^2} \int_{\mathcal{S}} z \mbox{e}^{i \kappa_x x} \mbox{e}^{i \kappa_z z}   \mbox{d}x \mbox{d}z \right) \mbox{d}\kappa_x \mbox{d}\kappa_z 
    \\
    =& \iint_{\mathbb{R}^2} \widehat{\tau}_{w,x,0}  \Psi^x_R  \, \mbox{d}\kappa_x \mbox{d}\kappa_z,
    \label{eq:tf_spectral}
\end{align}
where $\mathcal{S}$ is the disc surface, $\widehat{\tau}_{w,x,0}$ is the spatial Fourier transform of $\tau_{w,x,0}(x,z)$, and $\mathcal{F}^{-1}_{xz}\{\cdot\}$ is its inverse.
The function $\Psi^x_R(\kappa_x, \kappa_z)$ simplifies to:
\begin{equation}
    \Psi^x_R(\kappa_x, \kappa_z) = R^3 \Psi(R \kappa_x, R \kappa_z), 
\label{eq:psi_ss}
\end{equation}
where $\Psi$ is
\begin{equation}
    \Psi(\kappa_1, \kappa_2) = \frac{1}{2 \pi^2 \kappa_1} \int^1_{-1} z 
    \sin\left(\kappa_1 \sqrt{1 - z^2}\right) \mbox{e}^{i \kappa_2 z} \mbox{d} z.
\label{eq:psi_def}
\end{equation}

We then study the magnitudes $\vert \widehat{\tau}_{w,x,0} \vert$ and $\vert \Psi^x_R \vert$ as the largest contribution to Var($T_{f,0}^x$). The phase spectra $\Phi_\tau$ and $\Phi_\psi$ also contribute to Var($T_{f,0}^x$). The fixed-channel simulation data render $\Phi_\tau$ a function with a uniformly distributed random phase. The function $\Phi_\psi$ takes discrete values of either $-\pi/2$ or $\pi/2$, the sign being dictated by the same radial period observed for the oscillations of $\vert\Psi^x_R\vert$.

Figure \ref{fig:reduced_models}b shows the magnitude $\vert \Psi^x_R \vert$ of the spectral filter function, calculated by solving the integral in \eqref{eq:psi_def} numerically. The axes use the reciprocal of the wavelength (i.e., $1/\lambda_x$=$\kappa_x / 2\pi$) normalized with the disc diameter to visualize the filtering behaviour of $\Psi^x_R$ relative to the disc size. As the function $\vert \Psi^x_R \vert$ is even with respect to both wavenumbers, only the first quadrant is shown. A series of regularly-spaced oscillations can be observed, centred at the origin and decreasing in intensity at higher values of the diameter-wavelength ratio. The period of these oscillations, calculated along a radial direction centred at the origin, is $ D / \lambda$=1 where $\lambda = \sqrt{\lambda^2_x+\lambda^2_z}$, which means that $\vert \Psi^x_R \vert$ privileges spanwise modes, gradually decreasing at low spanwise wavenumbers and vanishing completely for purely streamwise modes. The maximum of $\vert \Psi^x_R \vert$ is determined to be a purely spanwise mode such that $D/\lambda_z = 0.732$. The largest individual contribution to the torque is given by the spanwise mode of $\tau_{w,x,0}$ whose wavelength is 1.37 times the diameter.

When observed in the absolute coordinates $1/\lambda_x$ and $1/\lambda_z$, the map of figure \ref{fig:reduced_models}b corresponds to the case $D=1$. When the disc size increases for $D>1$, the coordinates scale linearly with $D$, which means that the maximum of $\vert\Psi^x_R\vert$ moves closer to $\kappa_z=0$ and the radial wavelength of the oscillations decreases linearly, visually generating a ``shrinking'' effect centred at the origin. Similarly, when $D<1$, $\vert \Psi^x_R \vert$ undergoes an ``expansion'' according to the same proportionality of the coordinates to $D$. The other effect of varying the disc size is that $\vert \Psi^x_R \vert$ is uniformly amplified by a factor of $R^3$, according to \eqref{eq:psi_ss}. Larger discs therefore produce a very intense, localized filtering near the origin, while smaller discs have a weaker, spread-out filtering across a broad range of wavenumbers.

The PSD of the streamwise wall-shear stress reveals where the contributions to $\vert\widehat{\tau}_{w,x,0}\vert$ are located on average (not shown). It is found that, as $\vert\Psi^x_R\vert$ decreases and spreads out for increasingly large $D$, the  entire shear-stress PSD attenuates, leading to smaller torque values. Conversely, for sufficiently small $D$, $\Psi^x_R$ has an amplifying effect, with the most amplified modes being those of small streamwise wavenumbers, which correspond to the energy-containing region of the shear-stress PSD. Therefore, increasingly intense values of the torque occur as the disc size increases. This result qualitatively corresponds to that observed in figure \ref{fig:rms_Ds}, the precise dependence on $D$ being determined by the specific form of the shear-stress spectrum.

As discussed in \S\ref{sec:results-full}, the spanwise shear-stress torque $T^z_f$ is much smaller than the streamwise one. A relation analogous to \eqref{eq:tf_spectral} is derived for $T^z_f$:
\begin{equation}
    T_{f,0}^z(R) = \iint_{\mathbb{R}^2} \widehat{\tau_{w,z,0}}  \Psi^z_R  \, \mbox{d}\kappa_x \mbox{d}\kappa_z. 
\label{eq:tf_spectral_z}
\end{equation}
The difference between the spanwise filter function and the streamwise case is that $\Psi^z_R(\kappa_x, \kappa_z) = R^3 \Psi(R \kappa_z, R \kappa_x)$, i.e., the spanwise and the streamwise wavenumbers are exchanged. This result can be shown by substituting $z$ with $x$ in \eqref{eq:tf_spectral} and calculating the spatial integral. As implied by its definition, $\Psi^z_R$ privileges streamwise modes of $\widehat{\tau_{w,z,0}}$ (as opposed to spanwise for $\Psi^z_R$) and its maximum amplification happens for the purely streamwise mode such that $ D / \lambda_x \approx 0.732 $.

The variance of $T_{f,0}$ is calculated by evaluating the variance of the right-hand side of \eqref{eq:tf_spectral} and \eqref{eq:tf_spectral_z} over the ensemble of uncorrelated shear-stress fields produced from the numerical simulation of the fixed-wall channel. After taking the square root, the resulting curve,  measured directly from the spatial wall shear-stress fields through \eqref{eq:t_f} and shown in figure \ref{fig:reduced_models}b alongside $T_{f,0,\mbox{rms}}(D)$, shows excellent quantitative agreement with the numerical data.

It is conjectured that the change of slope occurring at around $D^+$=100 in figure \ref{fig:rms_Ds} originates from the scaling behaviour of the filter-function kernel $\Psi^x_R$ with respect to $D$. According to \eqref{eq:psi_ss} and the visualization shown in figure \ref{fig:reduced_models}b, $\Psi^x_R$ has two possible mechanisms for amplifying the wall-shear-stress modes. First, its amplitude grows as $D^3$ and, second, its maxima move relative to the wall-shear-stress spectrum maximum because the coordinates scale with $D$. While the first mechanism always results in an uniformly increasing amplification, the effect of the second mechanism is significant only when the $\Psi^x_R$ maximum overlaps the region of maximum energy concentration of the wall-shear stress. On the contrary, if the $\Psi^x_R$ maximum overlaps modes for which the wall-shear-stress PSD is zero, no local amplification takes place and those modes do not contribute to the integral \eqref{eq:tf_spectral}. Therefore, as $D$ increases and the $\Psi^x_R$ maximum moves closer to the origin, initially a regime is expected where both the first and the second mechanisms work in favour of the amplification because the kernel maximum moves progressively closer to the PSD maximum, until the two overlap.
Thereafter, a regime of reduced amplification rate, i.e., a gentler slope of the $T_{f,\mbox{rms}}(D)$ curve, is expected because the two maxima move apart. The diameter $D_c$ at which the maximum of $\vert\Psi^x_R\vert$ and the maximum of $S_{\tau\tau}$ overlap can be used as an estimate of the slope-changing diameter. It is found that the maximum of $S_{\tau\tau}$ is located at around $1/\lambda_z = 1.2$, which implies that $D_c$=0.732/1.2=0.61 or $D_c^+ \approx 110$ in viscous units. This value matches the slope-changing point of $T_{f,0,\mbox{rms}}(D)$ well, qualitatively endorsing the conjectured mechanism.

\end{document}